\documentclass[letter]{aa}  

\usepackage{graphicx}
\usepackage{txfonts}
\graphicspath{ {./figures/} }
\newcommand{\Poincare}{Poincar\'{e}}
\newcommand{\arcsecyr}{$^{\prime\prime} \,{\textrm{yr}}^{-1}$}
\newcommand{\arcsecyrtext}{${\textrm{arcsec yr}}^{-1}$}

\let\vec\boldsymbol

\newcommand*\dif{\mathop{}\!\mathrm{d}\!}

\renewcommand{\eqref}[1]{(\ref{#1})}
\newcommand{\suppref}[1]{(Appendix \ref{#1})}
\def\E{\mathrm{E}}
\def\inc{\mathcal{I}}
\def\F{\mathrm{F}}

\newcommand{\nomsol}{$\mathcal{S}$}



\newcommand{\subscript}[2]{#1_{\textnormal{\scriptsize #2}}}

\newcommand{\Hsec}{\widehat{H}}
\newcommand{\Hiss}{\mathcal{H}}

\newcommand{\out}[1]{\subscript{#1}{o}}

\newcommand{\vecomegaLL}{\subscript{\vec{\omega}}{LL}}

\newcommand{\vecu}{\vec{u}}
\newcommand{\vecv}{\vec{v}}

\newcommand{\Huv}{\mathbb{H}}

\newcommand{\Chi}{X}
\newcommand{\vecChi}{\vec{\Chi}}
\newcommand{\vecchi}{\vec{\chi}}
\newcommand{\vecPsi}{\vec{\Psi}}
\newcommand{\vecpsi}{\vec{\psi}}
\newcommand{\vecI}{\vec{I}}
\newcommand{\vectheta}{\vec{\theta}}
\newcommand{\veck}{\vec{k}}
\newcommand{\vecl}{\vec{\ell}}

\newcommand{\vecomegamean}{\overline{\vec{\omega}}}
\newcommand{\vecomegaout}{\out{\vec{\omega}}}

\newcommand{\fourcoeff}[3]{\widetilde{\Huv}_{#1}^{\vec{#2},\vec{#3}}}
\newcommand{\Ham}[1]{\Huv_{#1}}
\newcommand{\Hamstar}[1]{\Huv_{#1}^\star}
\newcommand{\lieham}{\Huv^\prime}

\newcommand{\lieI}{\vecI^\prime}
\newcommand{\lietheta}{\vectheta^\prime}
\newcommand{\liePhi}{\vec{\Phi}^\prime}
\newcommand{\liephi}{\vec{\phi}^\prime}
\newcommand{\lieomega}{\vec{\omega}^\prime}
\newcommand{\lieomegamean}{\vecomegamean^\prime}
\newcommand{\lieu}{\vecu^\prime}
\newcommand{\liev}{\vecv^\prime}

\newcommand{\liefourcoeff}[2]{\widetilde{\Huv}_{2n}^{\prime \, \vec{#1},\vec{#2}}}

\newcommand{\vecp}{\vec{p}}
\newcommand{\vecvarphi}{\vec{\varphi}}
\newcommand{\matM}{\mathbb{M}}
\newcommand{\vece}{\vec{e}}

\newcommand{\conj}[1]{\overline{#1}}

\def\F{\mathcal{F}}
\newcommand{\floor}[1]{\lfloor #1 \rfloor}

\def\omegaell{\omega_\mathrm{ell}}
\def\omegahyp{\omega_\mathrm{hyp}}
\def\halfwidth{\Delta\omega}

\def\timeres{\tau_\mathrm{res}}
\def\timelibr{\tau_\mathrm{libr}}

\begin{document}

\title{The origin of chaos in the Solar System through computer algebra}
\titlerunning{Origin of chaos in the Solar System}

\author{Federico Mogavero \and Jacques Laskar}
\authorrunning{F. Mogavero \and J. Laskar}

\institute{IMCCE, CNRS, Observatoire de Paris, Universit\'{e} PSL, Sorbonne Universit\'{e}, 
77 Avenue Denfert-Rochereau, 75014 Paris, France\\
\email{federico.mogavero@obspm.fr}}

\date{Received ; accepted }

 
\abstract{
The discovery of the chaotic motion of the planets in the Solar System dates back more 
than 30 years. Still, no analytical theory has satisfactorily addressed the origin 
of chaos so far. Implementing canonical perturbation theory in the computer algebra system TRIP, 
we systematically retrieve the secular resonances at work along the orbital solution of 
a forced long-term dynamics of the inner planets. 
We compare the time statistic of their half-widths to the ensemble distribution of the 
maximum Lyapunov exponent and establish dynamical sources of chaos in an unbiased way. 
New resonances are predicted by the theory and checked against direct integrations of 
the Solar System. The image of an entangled dynamics of the inner planets emerges. 
}


\maketitle
%

\section{Introduction}
The chaotic long-term behaviour of the planetary orbits in the inner Solar System (ISS)
emerged when the numerical integration of analytically averaged equations
of motion revealed a maximum Lyapunov exponent (MLE) of about (5 million years)$^{-1}$ 
\citep{Laskar1985,Laskar1989}. Previously, the existence of secular resonances 
among the precession frequencies of planet perihelia and nodes had been shown to generate
small divisors, which prevent the representation of the orbits as quasi-periodic series 
\citep{Laskar1984,Laskar1988}.
Investigating the origin of chaos, J.L. measured the libration period of
the Fourier harmonic $\theta_{2:1} = 2(g_3-g_4)-(s_3-s_4)$, which involves the fundamental frequencies 
of Earth and Mars, to be about 4.6 Myr \citep{Laskar1990}. He then proposed the 
libration--circulation transitions of the corresponding argument as a source 
of the observed MLE. Evidence of a large chaotic zone bridging 
the resonances $\theta_{2:1}$ and $\theta_{1:1} = (g_3-g_4)-(s_3-s_4)$ was later given \citep{Laskar1992}. 
Nevertheless, when the integration of the full equations of motion confirmed the MLE of 
the averaged ones, the chaotic nature of the $\theta_{2:1}$ dynamics 
was questioned \citep{Sussman1992}. As a consequence, a claim remains 
in literature that an undisputed dynamical mechanism for the observed chaos 
is missing \citep{Lecar2001,Murray2001,Hayes2007}. 
In the meantime, the alternating librations of $\theta_{2:1}$ and $\theta_{1:1}$
have been confirmed by direct integrations \citep{Laskar2004} and supported by geological 
records \citep{Ma2017,Olsen2019,Zeebe2019}. 

The high-dimensional dynamics of the inner planets probably discouraged systematic 
analytical studies of its resonant structure. A couple of analyses have focused 
on the long-term motion of Mercury, by freezing all the other planets on quasi-periodic 
orbits \citep{Lithwick2011,Batygin2015}, but this simplification leads to predictions 
that conflict with the findings of realistic models \citep{Mogavero2021}. 
Nevertheless, a Trojan horse against the curse of dimensionality affecting the ISS dynamics 
is offered by computer algebra, which allows the formal manipulation of the analytical 
series of celestial mechanics and the implementation of canonical perturbation theory in 
particular. Computer algebra has produced some remarkable results, such as the reproduction 
of Delaunay's monumental lunar theory \citep{Deprit1970} and the application 
of the Kolmogorov–Arnold–Moser theory to the three-body problem \citep{Robutel1995,Locatelli2000}, in addition to the 
demonstration of the chaotic behaviour of the Solar System itself \citep{Laskar1985,Laskar1989}. 
Still, its use in celestial mechanics may seem limited given its potential. 

We have recently proposed a forced model of the long-term dynamics in the ISS \citep{Mogavero2021}. 
It allows the secular phase space of the Solar System to be restricted in a consistent way  
to the eight degrees of freedom (DOFs) dominated by the inner planets. 
In this study we employ the computer algebra system TRIP \citep{Gastineau2011,TRIP} to carry out 
an unbiased analysis of the Fourier harmonics that constitute its Hamiltonian \suppref{supp:trip}.  

\section{Forced secular inner Solar System}
The long-term dynamics of the Solar System planets essentially consists of the slow 
precession of their perihelia and nodes, driven by secular, that is, orbit-averaged, 
gravitational interactions \citep{Laskar1990,Laskar2004}. The precession frequencies of the 
outer planet orbits are practically constant over billions of years when compared to those of the ISS
\citep{Laskar1990,Laskar2004,Hoang2021}. Built on these facts, the model of forced secular ISS 
consists in predetermining a quasi-periodic secular solution for the giant planets, 
with the inner ones moving in the resulting time-dependent gravitational potential \citep{Mogavero2021}. 
The quasi-periodic form of the giant planet orbits is established through frequency analysis 
\citep{Laskar2005} of the orbital solution of a comprehensive model of 
the Solar System \citep{Laskar2004}.  The low planetary masses and the absence of 
strong mean-motion resonances in the ISS allow us to simply consider 
first-order secular averaging of the $N$-body Hamiltonian. This corresponds to Gauss's 
dynamics of Keplerian rings \citep{Gauss1818}, which we correct for the leading secular 
contribution of general relativity. The pertinence of our model has been 
thoroughly demonstrated \citep{Mogavero2021}. It matches reference orbital solutions of the 
Solar System over timescales shorter than or comparable to the Lyapunov time, with an average 
discrepancy in the fundamental frequencies of only a few hundredths of an arcsecond per year 
over the next 20 Myr. Moreover, it correctly reproduces the MLE and the statistics of 
the high eccentricities of Mercury over the next 5 billion years (Gyr). 

\paragraph{Dynamical model.}
The secular Hamiltonian of the Solar System planets at first order in planetary masses and corrected 
for the leading contribution of general relativity reads \citep{Mogavero2021} 
\begin{equation}
\label{eq:ham_sec}
\Hsec = - \sum_{k=1}^8 G \frac{m_0 m_k}{a_k} 
\left[ \sum_{\ell=1}^{k-1} \frac{m_\ell}{m_0} \left< \frac{a_k}{\| \vec{r}_k - \vec{r}_\ell \|} \right> 
+ 3 \frac{G m_0}{c^2 a_k} \frac{1}{\sqrt{1-e_k^2}} \right]
.\end{equation}
The planets are indexed in order of increasing semi-major axis, $(m_k)_{k=1}^8$ are the planetary masses,
$a_k$ and $e_k$ are the (secular) semi-major axes and eccentricities, respectively, $G$ is the 
gravitational constant, and $c$ is the speed of light. The vectors $\vec{r}_k$ are the planet heliocentric positions, 
and the bracket operator represents the averaging over the mean longitudes of the planets $\lambda_k$, 
which results from the suppression of the non-resonant Fourier harmonics of the $N$-body Hamiltonian at first order 
in planetary masses \citep{Mogavero2021}. The semi-major axes $a_k$ are constants of motion in the secular dynamics. 
A suitable set of canonically conjugate momentum-coordinate pairs of variables for the secular dynamics are the \Poincare{} 
rectangular coordinates in complex form, $(x_k, -j \bar{x}_k; $ $y_k, -j \bar{y}_k)_{k=1}^8$, with 
\begin{equation}
\label{eq:poincare_vars}
\begin{aligned}
&x_k = \sqrt{\Lambda_k} \sqrt{1 - \sqrt{1- e_k^2}} \, \E^{j \varpi_k}, \\
&y_k = \sqrt{2 \Lambda_k} \left(1- e_k^2\right)^{\frac{1}{4}} \sin(\inc_k/2) \, \E^{j \Omega_k}, 
\end{aligned}
\end{equation}
where $\Lambda_k = \mu_k [G(m_0 + m_k) a_k]^{1/2}$, $m_0$ and $\mu_k = m_0 m_k / (m_0 + m_k)$ are the Sun 
mass and the reduced masses of the planets, respectively, $\inc_k$ are the planet inclinations, $\varpi_k$ are the 
longitudes of the perihelia, and $\Omega_k$ are the longitudes of the nodes \citep{Poincare1896,Laskar1991,Laskar1995}. 

The model of forced ISS consists of the choice of an explicit quasi-periodic time dependence for the 
orbits of the outer planets \citep{Mogavero2021}, 
\begin{equation}
\label{eq:qp_decomposition}
x_k(t) = \sum_{\ell=1}^{M_k} \tilde{x}_{k\ell} \, 
\E^{j \vec{m}_{k\ell} \cdot \vec{\phi}(t)}, \quad
y_k(t) = \sum_{\ell=1}^{N_k} \tilde{y}_{k\ell} \, 
\E^{j \vec{n}_{k\ell} \cdot \vec{\phi}(t)} 
,\end{equation}
for $k \in \{5,6,7,8\}$, where $t$ denotes the time, $\tilde{x}_{k\ell}$ and $\tilde{y}_{k\ell}$ are complex amplitudes, 
$\vec{m}_{k\ell}$ and $\vec{n}_{k\ell}$ are integer vectors, and $\vec{\phi}(t) = \out{\vec{\omega}} t$, with 
$\out{\vec{\omega}} = (g_5,g_6,g_7,g_8,s_6,s_7,s_8)$ representing the septuple of the constant fundamental frequencies of the outer orbits \citep{Laskar1990}. 
Gauss's dynamics of the forced ISS is obtained by substituting this predetermined time dependence into Eq.~\eqref{eq:ham_sec}, 
\begin{equation}
\label{eq:ham_sec_inn}
\Hiss[(x_k,y_k)_{k=1}^4,t] = 
\Hsec[(x_k,y_k)_{k=1}^4, (x_k = x_k(t), y_k = y_k(t))_{k=5}^8] 
.\end{equation}
The resulting Hamiltonian system consists of two DOFs for each inner planet, corresponding to the 
$x_k$ and $y_k$ variables, respectively. The forced secular ISS is thus characterised by eight DOFs. 
As a result of the forcing from the outer planets, its orbital solutions live in a 16-dimensional phase space 
since no trivial integrals of motion, such as the total energy or angular momentum, exist. 

The development of the two-body perturbing function \citep{Laskar1995}, 
when implemented in TRIP, allows the Hamiltonian $\Hiss$ to be systematically expanded 
by exploiting the low eccentricities and inclinations of the planets \citep{Mogavero2021}. 
This development provides truncated Hamiltonians $\Hiss_{2n}$ that are multivariate 
polynomials of total degree $2n$ in the \Poincare{} variables of the inner planets \suppref{supp:model}. 
At the lowest degree, $\Hiss_2$ produces a forced Laplace-Lagrange (LL) dynamics, 
which can be analytically integrated by introducing complex proper mode variables, $(u_k, v_k)_{k=1}^4$. 
By introducing action-angle variables through $u_k = \sqrt{\Chi_k} \E^{-j \chi_k}$ and $v_k = \sqrt{\Psi_k} \E^{-j \psi_k}$, 
the truncated Hamiltonians expressed in the proper modes can be expanded as finite Fourier series, 
\begin{equation}
\label{eq:hamiltonian}
\Huv_{2n}(\vecI,\vectheta,t) = 
\sum_{\vec{k} \in \mathbb{Z}^8}
\sum_{\vec{\ell} \in \mathbb{Z}^7} 
\fourcoeff{2n}{k}{\ell} (\vecI)
\E^{j \left( \vec{k} \cdot \vec{\theta} + \vec{\ell} \cdot \out{\vec{\omega}} t \right)} 
,\end{equation}
where $\vecI = (\vecChi,\vecPsi)$ and $\vectheta = (\vecchi,\vecpsi)$ are the eight-dimensional vectors of 
the action and angle variables, respectively, and $(\veck, \vecl)$ is the wave vector of a given harmonic. 
There are 2\,748 harmonics with a non-null amplitude $\fourcoeff{2n}{k}{\ell}$ at degree four 
and more than ten million at degree ten \citep{Mogavero2021}. 

\section{Maximum Lyapunov exponent}
\begin{figure}
\includegraphics[width=\columnwidth]{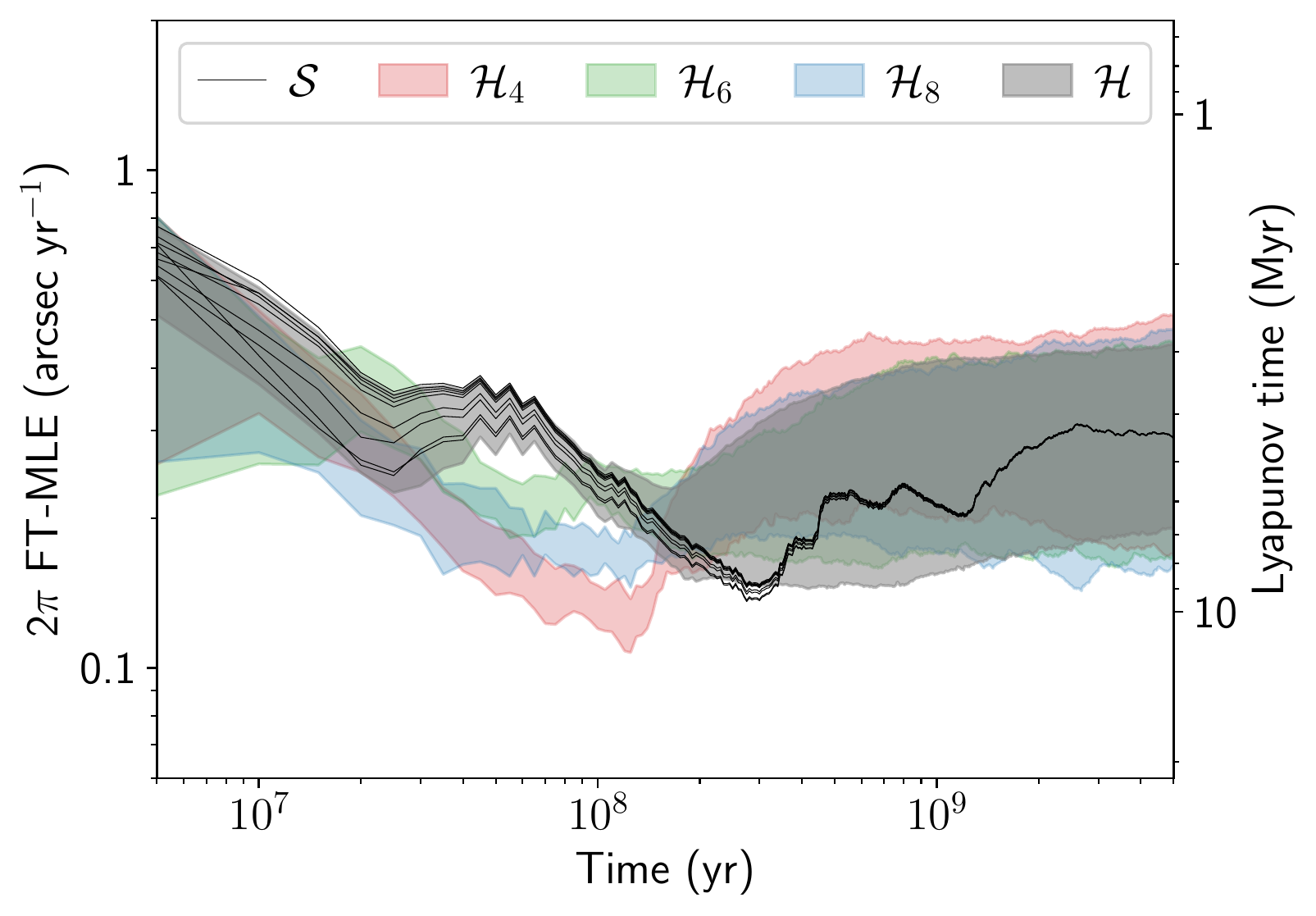}
\caption{Finite-time MLE of the forced secular ISS over 5 Gyr for different degrees of truncation of the Hamiltonian and the corresponding Lyapunov time FT-MLE$^{-1}$. 
The bands represent the [5$^{\textrm{th}}$, 95$^{\textrm{th}}$] percentile range of the PDF estimated 
from ensembles of 128 ($\Hiss_{2n}$) and 1148 ($\Hiss$) stable orbital solutions with very close initial conditions. 
The black lines denote the nominal solution \nomsol{} for nine different initial tangent vectors.}
\label{fig:lyap}
\end{figure}

Computing the MLE is fundamental to the determination of the origin of chaos, 
as its value can be compared to the half-width of the leading resonant harmonics 
of the Hamiltonian, which constitute the dynamical sources of chaoticity \citep{Chirikov1979}. 
The non-null probability of unstable orbital evolutions in the ISS \citep{Laskar2009} makes 
the definition of the MLE as an infinite-time limit \citep{Oseledec1968,Benettin1980} not pertinent. 
We numerically compute a finite-time MLE (FT-MLE) employing the standard algorithm of \citet{Benettin1980} 
(faster chaoticity detectors have been developed starting from the fast Lyapunov indicator of \citealt{Froeschle1997}). 
The FT-MLE is time-asymptotically a stochastic function of the initial conditions of the system, and its computation acquires full physical 
significance for an ensemble of orbital solutions \citep{Mogavero2021}. Manipulation of the truncated Hamiltonians $\Hiss_{2n}$ 
in TRIP allows us to systematically derive the equations of motion and the corresponding 
variational equations, which we integrate through an Adams PECE method of order 12 and a  
timestep of 250 years. Figure \ref{fig:lyap} shows the FT-MLE expressed as an angular frequency 
over the next 5 Gyr for different degrees of truncation of the Hamiltonian. 
In each case, the FT-MLE is computed for 128 stable (i.e. non-collisional) solutions, with initial conditions very close to 
the nominal values of Gauss's dynamics and for different initial tangent vectors \suppref{supp:nomsol}.  The figure shows the [5$^{\textrm{th}}$, 95$^{\textrm{th}}$] percentile range of 
the probability distribution function (PDF) of the FT-MLE estimated from each ensemble of solutions. 
We also report the PDF from the full Hamiltonian $\Hiss$, computed in \citet{Mogavero2021}, 
along with the FT-MLE of its nominal solution \nomsol{} for nine different initial tangent vectors, 
to manifest its asymptotic behaviour. In a few hundred million years, each FT-MLE becomes independent 
of the initial tangent vector, the renormalisation time, and the norm chosen for the phase-space vectors. 
Its distribution only reveals the intricate dependence on the initial position of the system in the 
phase space. 
At 5 Gyr, the FT-MLE roughly ranges from 0.15 to 0.5\arcsecyr{} with a 90\% probability. 
Figure \ref{fig:lyap} shows that the truncated Hamiltonians $\Hiss_{2n}$ reproduce the asymptotic distribution 
of the FT-MLE of the full dynamics $\Hiss$, even at the lowest degree. It suggests, in particular, that 
the dynamical interaction of the Fourier harmonics at degree four constitutes the primary source of the 
observed FT-MLE. 

\section{Leading harmonics}
Significant insight into the ISS dynamics is provided by the knowledge of the leading harmonics of the Hamiltonian, that is, 
those that drive the system trajectory in the action space the most, without necessarily being resonant. 
The action variables in turn  control the essential long-term variation in the frequencies of motion 
(and therefore the activation of a specific web of resonances) through the mean frequencies 
$\vecomegamean_{2n}(\vecI) = \partial \fourcoeff{2n}{0}{0} / \partial \vecI$ that derive from the integrable part 
$\fourcoeff{2n}{0}{0}$ of the Hamiltonian in Eq.~\eqref{eq:hamiltonian}. 
The contribution of the harmonic $\vec{k},\vec{\ell}$ to the action vector $\vecI(t)$ is 
\begin{equation}
\label{eq:harmonic_ranking_1}
\Delta \vecI_{\vec{k},\vecl}(t) = 
2 \vec{k} \operatorname{Im} 
\int_0^t 
dt^\prime \ \fourcoeff{2n}{k}{l}(\vecI(t^\prime)) \E^{j \left( \vec{k} \cdot \vec{\theta}(t^\prime) + \vecl \cdot \out{\vec{\omega}} t^\prime \right)} 
.\end{equation}
We ranked the 69\,339 harmonics of $\Huv_6$ according to the time median of the relative Euclidean norm 
$\delta\vecI_{\vec{k},\vecl}(t) = \Vert \Delta \vecI_{\vec{k},\vecl}(t) \Vert / \left\Vert \vecI(t) \right\Vert$ 
along the nominal solution \nomsol{} of Gauss's dynamics, spanning 5 Gyr \suppref{supp:leading_harm}. 
We report the first 30 harmonics in Table~\ref{tab:leading_harmonics}. As usual, each harmonic is identified by the 
corresponding combination of frequency labels $(g_i, s_i)_{i=1}^8$ \citep{Laskar1990}. Given the absence of 
harmonics of order six, Table~\ref{tab:leading_harmonics} confirms the leading role of those of order two 
and four, which enter the truncated Hamiltonian $\Huv_{2n}$ at degree four. The harmonics $\theta_{1:1}$ and 
$\sigma_{1:1} = (g_1 - g_5) - (s_1 - s_2)$ appear among the very leading terms, confirming their dynamical relevance, as suggested in 
previous studies \citep{Laskar1990,Laskar1992,Lithwick2011,Boue2012,Batygin2015}. Among the top terms, there are harmonics that couple more extensively the DOFs of the inner 
planets, with the remarkable examples of $(g_1 - g_4) + (s_1 - s_4),$ associating the proper modes of Mercury and Mars and $(s_1 - s_2) + (s_3 - s_4),$ 
which concatenates all the inclination DOFs. A non-negligible role of the Saturn-dominated eccentricity mode $g_6$ also emerges. 
Even though these harmonics are not all necessarily resonant, they suggest that the dynamical entanglement 
of all the DOFs is significant along the nominal solution \nomsol{}. This consideration is supported by the study of partial Hamiltonians 
constructed from a limited number of Fourier harmonics \suppref{supp:partial_harm}. As Fig. \ref{fig:lyap_partials} shows, 
while the harmonics in Table~\ref{tab:leading_harmonics} allow the asymptotic FT-MLE of Fig.~\ref{fig:lyap} to be robustly reproduced, 
simplified Hamiltonians based on the selection of specific DOFs may provide inconsistent predictions, 
with too low values down to a non-chaotic dynamics. This notably indicates that the resonant nature of 
a Fourier harmonic should only be established along the orbital solution of a realistic Hamiltonian. 

\begin{table}
\caption{Leading Fourier harmonics of $\Huv_6$ along the nominal solution \nomsol{} spanning 5 Gyr.} 
\label{tab:leading_harmonics} 
\centering
\begin{tabular}{r r c | r r c} 
\hline\hline
\rule{0pt}{2.ex}
$i$ & Harmonic $[\F_i]$ & $\delta\vecI_{\vec{k},\vecl}$ & & & \\ 
\hline
\rule{0pt}{2.1ex}
\input{files/harm_table_twocolumns.out}
\hline
\end{tabular}
\tablefoot{First 30 harmonics ranked according to the time median of their relative contribution 
$\delta\vecI_{\vec{k},\vecl}(t)$ to the action vector $\vecI(t)$. 
The 5$^{\mathrm{th}}$ and 95$^{\mathrm{th}}$ percentiles are reported as subscripts and superscripts, 
respectively.}
\end{table}

\begin{figure*}
\includegraphics[width=2\columnwidth]{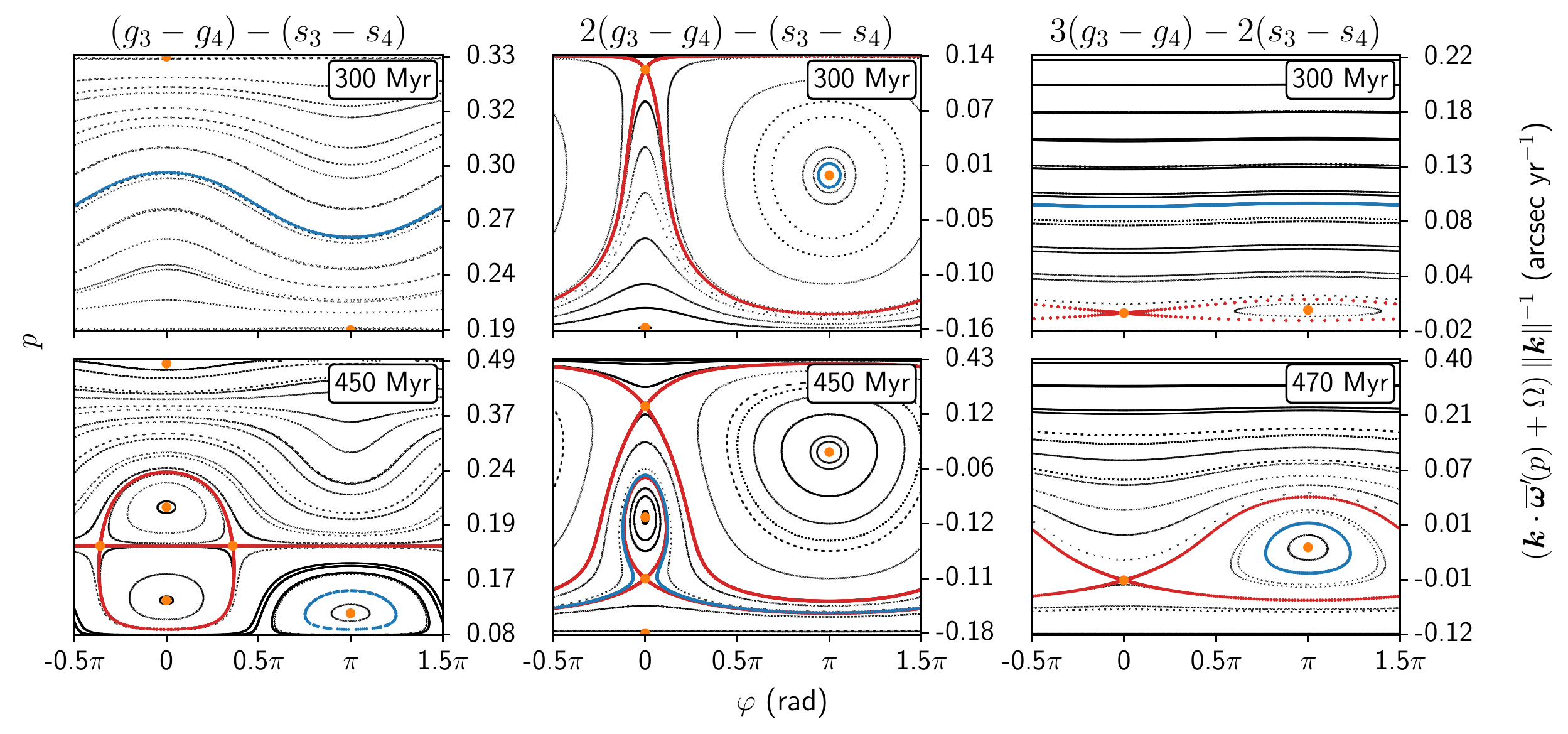}
\caption{Reduced phase spaces $(p,\varphi)$ of the harmonics $\theta_{1:1}$, $\theta_{2:1}$, and $\theta_{3:2}$ along the 
Lie-transformed nominal solution $\mathcal{S}_{10}^\prime(t)$ as deduced from the Lie-transformed Hamiltonian $\lieham_{10}$. 
The phase spaces correspond to the $t_0$ times indicated in the upper-right boxes. 
The black dots reproduce the level curves of the reduced Hamiltonian, and the orange dots 
represent its fixed points. The separatrices are shown by red dots, and the level curve corresponding to the 
resonant variables at time $t_0$ is shown in blue. The right-hand-side axes report the projection of the mean frequencies 
$\lieomegamean_{2n}(p) = \partial \liefourcoeff{0}{0} / \partial \lieI$ along the normal $\veck$ to the resonance plane.}
\label{fig:phase_space_1}
\end{figure*}

\section{Resonant harmonics}
Figure~\ref{fig:lyap} and Table~\ref{tab:leading_harmonics} suggest that the non-linear interaction of the Fourier 
harmonics at degree four constitutes the primary source of chaos. In the framework of canonical perturbation theory, 
we constructed a Lie transform to define a change of variables that eliminates the terms of degree four from the 
truncated Hamiltonian $\Huv_{2n}$ \suppref{supp:lie}. In the Lie-transformed Hamiltonian, $\lieham_{2n}$, which is in 
Birkhoff normal form to degree four, these terms are replaced with the chain of high-order harmonics that arise 
from their dynamical interaction. Very importantly, the aim of this procedure is not to set up successive analytical 
approximations of the dynamics, which would be a vain goal. We simply define new canonical variables  that let the 
interactions of the terms of degree four appear explicitly in the amplitudes of the Fourier harmonics at higher degrees. 

To reveal the resonant harmonics of the Hamiltonian $\lieham_{2n}$, we first retrieved those that present episodes 
of libration along the Lie-transformed nominal solution, $\mathcal{S}_{2n}^\prime$. 
Following \citet{Mogavero2021}, we defined time-dependent fundamental frequencies for the inner orbits $(g_i^\prime(t), s_i^\prime(t))_{i=1}^4$ \suppref{supp:librating_harm}. 
For each harmonic of $\lieham_{2n}$, we then evaluated the corresponding combination of frequencies along the nominal solution. 
When this combination becomes null at least once over the time span of the solution, we consider the harmonic to be librating. 
This procedure filters out a majority of the Fourier harmonics as they never librate. 

The appearance of libration episodes, and thus the existence of libration islands in the phase space, 
does not guarantee the resonant nature of a harmonic, as this is connected to the presence of 
both stable and unstable manifolds. Therefore, we employed the classic divide-et-impera approach 
of \citet{Chirikov1979} and considered a reduced Hamiltonian for each librating harmonic: 
\begin{equation}
\label{eq:reduced_hamiltonian}
\hslash_{2n}^{\veck,\vecl}(p,\varphi) = \liefourcoeff{0}{0}(\lieI(p)) + \Omega \, p 
+  2 \operatorname{Re} \{ \liefourcoeff{k}{\ell}(\lieI(p)) \, \E^{j \varphi} \} 
.\end{equation}
The resonant variables $p,\varphi$ are related to the Lie-transformed action-angle variables $\lieI,\lietheta$ 
by $\lieI(p) = \lieI_0 + p \veck$ and $\varphi = \veck \cdot \lietheta + \Omega t$, 
where we denote $\Omega = \vecl \cdot \vecomegaout$ \suppref{supp:reduced_ham}. 
We therefore inspected the fictitious one-DOF dynamics that would be generated if $\veck,\vecl$ were 
the only harmonic appearing in $\lieham_{2n}$. 
The topology of the reduced phase space $(p,\varphi)$ depends on seven integrals of motion, whose values relate 
to the position $\lieI_0$ of the system in the full-dimensional action space. For a given librating harmonic, 
our study considers the one-parameter family of the reduced phase spaces that arise at each point along the nominal solution 
$\mathcal{S}_{2n}^\prime(t) = (\lieI_{2n}(t), \lietheta_{2n}(t))$, that is, $\lieI_0 = \lieI_{2n}(t=t_0)$ \suppref{supp:reduced_ham}. 
This family of phase spaces is therefore spanned by the time $t_0$. As an example, Fig.~\ref{fig:phase_space_1} shows, for different $t_0$ 
times along $\mathcal{S}_{10}^\prime(t)$, the phase spaces of the harmonics $\theta_{1:1}$, $\theta_{2:1}$, and $\theta_{3:2}$, with $\theta_{m:n} = m(g_3-g_4)-n(s_3-s_4)$, 
as deduced from the Hamiltonian $\lieham_{10}$. The times were chosen across the first transition of $\theta_{2:1}$ from libration to circulation, 
which occurs at about 340 Myr along the nominal solution $\mathcal{S}_{10}^\prime$, and corresponds to the point where its FT-MLE starts 
to increase in Fig.~\ref{fig:lyap}. The figure shows the level curves of the reduced Hamiltonian reconstructed by numerical integration as well as its fixed points, computed semi-analytically with TRIP. For a resonant harmonic, 
separatrices emerge from the hyperbolic fixed points and enclose libration islands. 
The level curve corresponding to the value of the resonant variables at time $t_0$ along $\mathcal{S}_{10}^\prime(t)$ is also shown 
and used to define the temporary libration or rotation state of the harmonic. 

Figure \ref{fig:phase_space_1} shows how a resonant phase space can significantly differ from that of a simple pendulum, 
which is the universal model of non-linear resonance and is often invoked to perform analytical computations. 
While the pendulum approximation turns out to be well suited for the majority of the resonant harmonics, 
it does not apply in general to the leading resonances, which are our main interest. 
We therefore discarded this approximation and performed systematic algebraic manipulations with TRIP to retrieve 
the fixed points of the reduced phase spaces through the roots of a univariate polynomial in the action $p$ \suppref{supp:reduced_ham}. 
Relying on a polynomial solver \citep[Appendix~\ref{supp:reduced_ham},][]{Bini1996,Bini2000,Bini2014}, we established the resonant nature of each harmonic along the nominal solution 
in a numerically robust and efficient way. We retrieved, in particular, the characteristic frequencies of the motion $\omegaell$ 
and $\omegahyp$ around the elliptic and hyperbolic fixed points, respectively. These frequencies are related to the
square root of the eigenvalues of the variational matrix that governs the linear stability of the fixed points. 
Via a similar procedure, we retrieved the extrema of the action $p$ along a level curve of the 
reduced Hamiltonian, which are employed to determine the libration-rotation state of the harmonic and 
the resonance half-width, $\halfwidth$. Following \citet{Chirikov1979}, we considered the projection 
of the frequency vector $\lieomega = d\lietheta/dt$ along the normal to the resonance plane $\veck \cdot \lieomega + \Omega = 0$, 
that is, $\omega = \lVert\veck\rVert^{-1} \veck \cdot \lieomega$. The reduced dynamics in Eq.~\eqref{eq:reduced_hamiltonian} 
induces a variation in this projection equal to $\halfwidth = \lVert\veck\rVert^{-1} \Delta \dot{\varphi}$. 
Building on ideas behind frequency analysis \citep{Laskar1993}, we then defined the resonance half-width $\halfwidth$ 
from the variation in the rotational frequency of the angle $\varphi$ across the separatrix \suppref{supp:reduced_ham}. 
In the pendulum approximation we have, up to a constant factor close to one, 
\begin{equation}
\label{eq:chirikov_halfwidth}
\halfwidth \sim 2 \omegahyp \lVert\veck\rVert^{-1} 
,\end{equation}
which is the \citet{Chirikov1979} half-width (recall that $\omegaell = \omegahyp$ for the pendulum). 

\section{Origin of chaos}
To establish a meaningful connection between the resonant phase spaces and the observed FT-MLE, we 
assumed that the time distribution of physical observables along the nominal solution spanning 5 Gyr 
is representative of their ensemble distribution (that is, over a set of stable orbital solutions 
with very close initial conditions) at some large time of the order of billions of years. Based on this sort of 
finite-time ergodic assumption, we systematically retrieved for each librating harmonic of $\lieham_{2n}$ 
the samples of the fixed point frequencies $\omegahyp$, $\omegaell$ and of the resonance half-width 
$\halfwidth$ along the nominal solution $\mathcal{S}_{2n}^\prime$. Table~\ref{tab:resonant_harmonics} shows the 
first 30 resonant harmonics, along with their time statistic, as deduced from the truncation of the Lie transform 
at degree ten. We report for each frequency its time median as well as the 5$^{\textrm{th}}$ and 95$^{\textrm{th}}$ 
percentiles as subscript and superscript, respectively. The resonances are ranked according to 
their median half-width. To measure the overall dynamical impact of the terms, we denote as $\timeres$ the fraction 
of time a harmonic is resonant and as $\timelibr$ the fraction of libration states in the resonant case. 
Table~\ref{tab:resonant_harmonics} shows the harmonics that are resonant for more than 1\% of the time. 

\begin{table*}
\caption{First 30 resonant harmonics of $\lieham_{10}$ along the nominal solution $\mathcal{S}^\prime_{10}$ spanning 5 Gyr.} 
\label{tab:resonant_harmonics}
\centering
\begin{tabular}{r c r | c r | c r | c} 
\hline\hline
\rule{0pt}{2.1ex}
$i$ & & Fourier harmonic $[\F_i]$ & $\omegahyp$ & $\timeres$ & $\omegaell$ & $\timelibr$ & $\halfwidth$ \\ 
\hline
\rule{0pt}{2.1ex}
\input{files/resonance_table_H10_final.out}
\hline
\end{tabular}
\tablefoot{Time median of the fixed point frequencies $\omegahyp$, $\omegaell$ and of the resonance half-width $\halfwidth$ 
in \arcsecyrtext{}. The 5$^{\mathrm{th}}$ and 95$^{\mathrm{th}}$ percentiles are reported as subscripts and superscripts, 
respectively. The fraction of time the harmonic is resonant is $\timeres$, and $\timelibr$ represents the fraction of libration states 
in the resonant case. The resonances are ranked according to their median half-width.} 
\end{table*}

\begin{figure*}
\centering
\includegraphics[width=1.54\columnwidth]{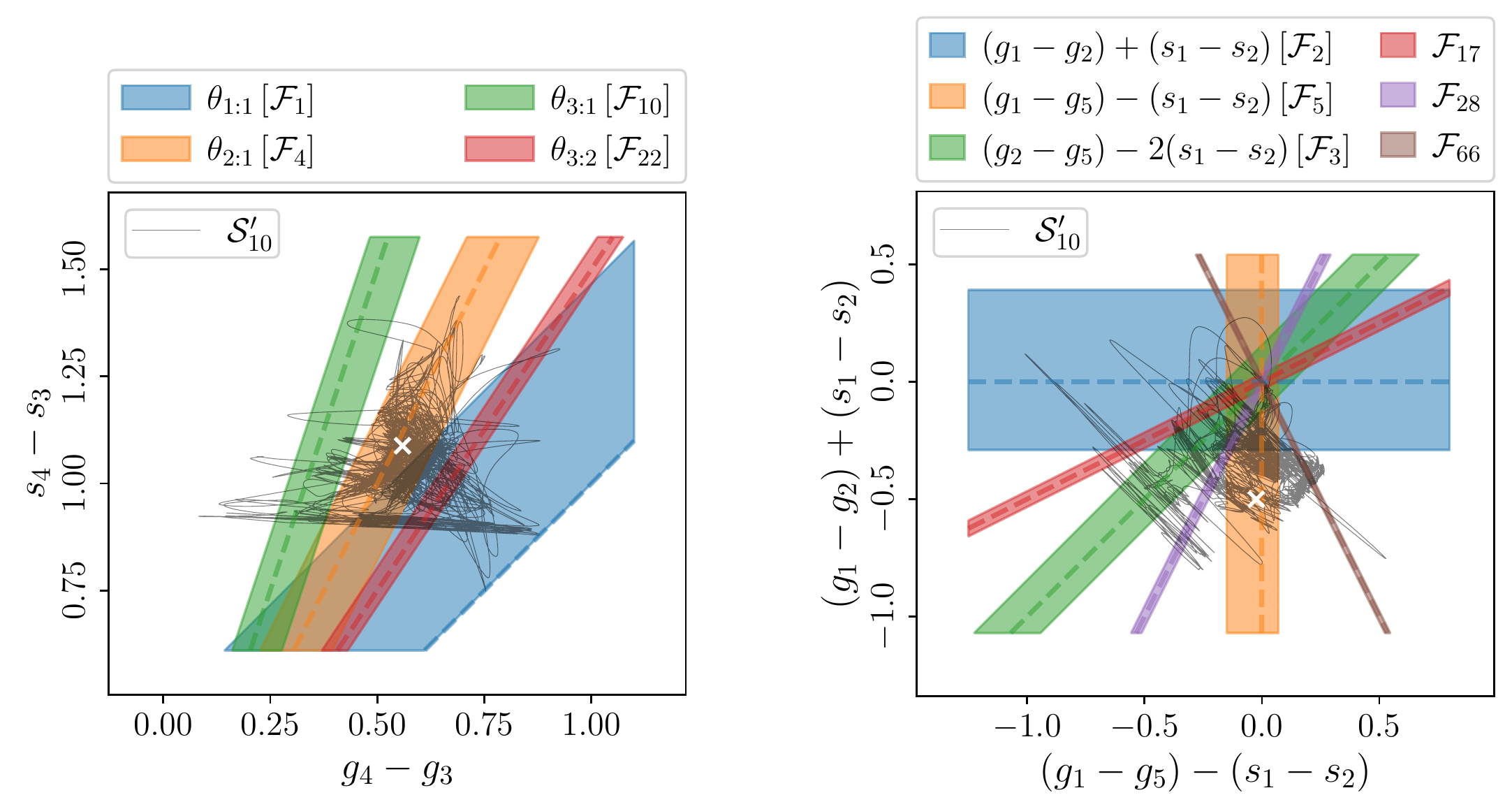}
\caption{Overlaps of secular resonances from Table~\ref{tab:resonant_harmonics}. Left panel: Starred resonances $\theta_{m:n} = m(g_3-g_4)-n(s_3-s_4)$ 
in the frequency subspace spanned by $g_4-g_3$ and $s_4-s_3$. Right panel: Daggered resonances as integer combinations of $(g_1-g_5)-(s_1-s_2)$ 
and $(g_1-g_2)+(s_1-s_2)$ (the harmonic $\F_{66} = 3g_1-g_2-2g_5-(s_1-s_2)$ is not shown in Table~\ref{tab:resonant_harmonics}). 
The dashed lines represent the resonance centres, and the coloured regions correspond to the time median of the signed half-widths. 
The nominal solution $\mathcal{S}^\prime_{10}$, which spans 5 Gyr, is shown after low-pass filtering with a cutoff frequency of $(\textrm{5 Myr})^{-1}$, 
the cross indicating the current position of the ISS. Frequencies are in \arcsecyrtext{}.}
\label{fig:overlaps}
\end{figure*}

It is perhaps hopeless to obtain, for high-dimensional dynamical systems, a precise formula connecting 
the FT-MLE to the half-width of the leading resonances or to their fixed point frequencies. Nevertheless, 
studies of the periodically forced pendulum \citep{Chirikov1979,Holman1996,Murray1997,Li2014} 
suggest that these quantities should differ only by a factor of order unity from one another, that is, 
\begin{equation}
\label{eq:MLE_halfwidth}
2\pi \, \textrm{FT-MLE} \sim \halfwidth
.\end{equation}
To fix ideas, we also assumed that Eq.~\eqref{eq:chirikov_halfwidth} holds beyond pendulum approximation. 
For the top resonances of Table~\ref{tab:resonant_harmonics}, the statistical intervals 
of both $\halfwidth$ and $\omegahyp$ extensively overlap the range spanned by the asymptotic FT-MLE 
of the nominal solution in Fig.~\ref{fig:lyap}. Moreover, they are significantly compatible with 
the long-term ensemble distribution of the FT-MLE. A statistical connection between the leading resonances of 
the Hamiltonian and the MLE thus emerges, even in the absence of relations more precise than Eqs.~\eqref{eq:MLE_halfwidth} 
and \eqref{eq:chirikov_halfwidth}. It indicates the dynamical sources of the observed chaos. 
One may appreciate the low number of resonances with $\halfwidth \gtrsim 0.1$\arcsecyr{},
which constitutes a sounding explanation for the large width of the FT-MLE distribution, while the dense set of 
terms with $\halfwidth < 0.1$\arcsecyr{} explains why the exponent never goes below this value. We point out 
the absence of resonances of order two. The harmonic $g_1-g_5$, in particular, is known to be involved in the 
very high eccentricities of Mercury \citep{Boue2012} and is not expected to become resonant along a stable solution. 

The resonances marked by a star in Table~\ref{tab:resonant_harmonics} only involve the proper modes 3 and 4, 
while those distinguished by a dagger exclusively concern the proper modes 1, 2, and 5. The first group of harmonics is 
composed of $\theta_{1:1}$, $\theta_{2:1}$, $\theta_{3:1}$, and $\theta_{3:2}$ and is associated with the 
chaotic zone proposed in \citet{Laskar1992}. The second group belongs to the frequency 
subspace spanned by the combinations $g_1-g_5$, $g_2-g_5$, and $s_1-s_2$ and includes 
the well-known resonance $\sigma_{1:1}$. We first point out that the libration frequencies of 0.28 
and 0.12\arcsecyr{} associated with $\theta_{2:1}$ and $\sigma_{1:1}$ in \citet{Laskar1990} 
are consistent with the corresponding statistics of the elliptic fixed point frequency $\omegaell$. 
We then show that each of these two sets of harmonics 
constitutes a source of chaoticity. The emergence of chaos from resonant one-DOF phase spaces, 
as in Fig.~\ref{fig:phase_space_1}, can be stated in two ways. 
As time changes and the system moves along its nominal trajectory, the phase-space region swept by the separatrices 
forms a chaotic zone. In an equivalent way, a chaotic zone results, at fixed time $t_0$, from 
separatrix-splitting when one restores the harmonics of $\lieham_{2n}$ that are suppressed in the reduced Hamiltonian. 
In any case, chaos derives from the interaction of resonant harmonics. Figure \ref{fig:overlaps} represents such an interaction 
in terms of resonance overlap \citep{Chirikov1979} for the groups of starred and daggered harmonics. It shows the resonance planes 
in the frequency space  along with asymmetric resonance layers, defined by the time median of the signed half-widths $\halfwidth^+$, $\halfwidth^-$ \suppref{supp:reduced_ham}. 
The reported half-widths are meaningful in the region visited by the nominal solution, which is also shown. 
The existence of the chaotic zone suggested in \citet{Laskar1992} is indeed supported by a 
statistically robust overlap. The resonance $\theta_{2:1}$, in particular, appears right in the centre of the resonance network, 
in agreement with its large $\timeres$ value. Its amplitude results to a large extent from the interaction of $\theta_{1:1}$ 
and $\theta_{1:0} = g_3-g_4$ at degree four (see Table~\ref{tab:leading_harmonics}). The resonance overlap in the $g_1-g_5$, $g_2-g_5$, $s_1-s_2$ subspace 
is shown in Fig.~\ref{fig:overlaps} in terms of integer combinations of $\sigma_{1:1}$ and $(g_1-g_2)+(s_1-s_2)$. It appears to be 
more relevant, at least for our nominal solution, than the restriction to the subspace spanned by $g_1-g_5$ and $s_1-s_2$ investigated in previous 
studies \citep{Lithwick2011,Batygin2015}. A relevant dynamical role of the proper mode $g_2$, hinted at in \citet{Lithwick2011} and \citet{Batygin2015}, 
is confirmed. Figure \ref{fig:overlaps} also suggests a long-term diffusion mainly perpendicular to the resonance 
planes, while Arnold's diffusion along them seems negligible \citep{Laskar1993}, at least for the present regular (i.e. non-collisional) solution. 

Right after the top resonances, harmonics with $\halfwidth < 0.1$\arcsecyr{} significantly couple all the proper modes, 
confirming the implications of Table~\ref{tab:leading_harmonics}. 
Coupling resonances such as $(g_2-g_4)+(s_2-s_4)$, $(g_1-g_3)+(s_2-s_3)$ and $(g_1-g_4)+(s_1-s_4)$, along with high-order ones 
resulting from their interaction with leading harmonics, strongly suggest that resonance overlap takes place extensively 
in the full-dimensional frequency space. Even though we have isolated low-dimensional sources of chaoticity, the large-scale chaos 
in the ISS is probably best understood as a high-dimensional phenomenon. 

\begin{figure}
\includegraphics[width=\columnwidth]{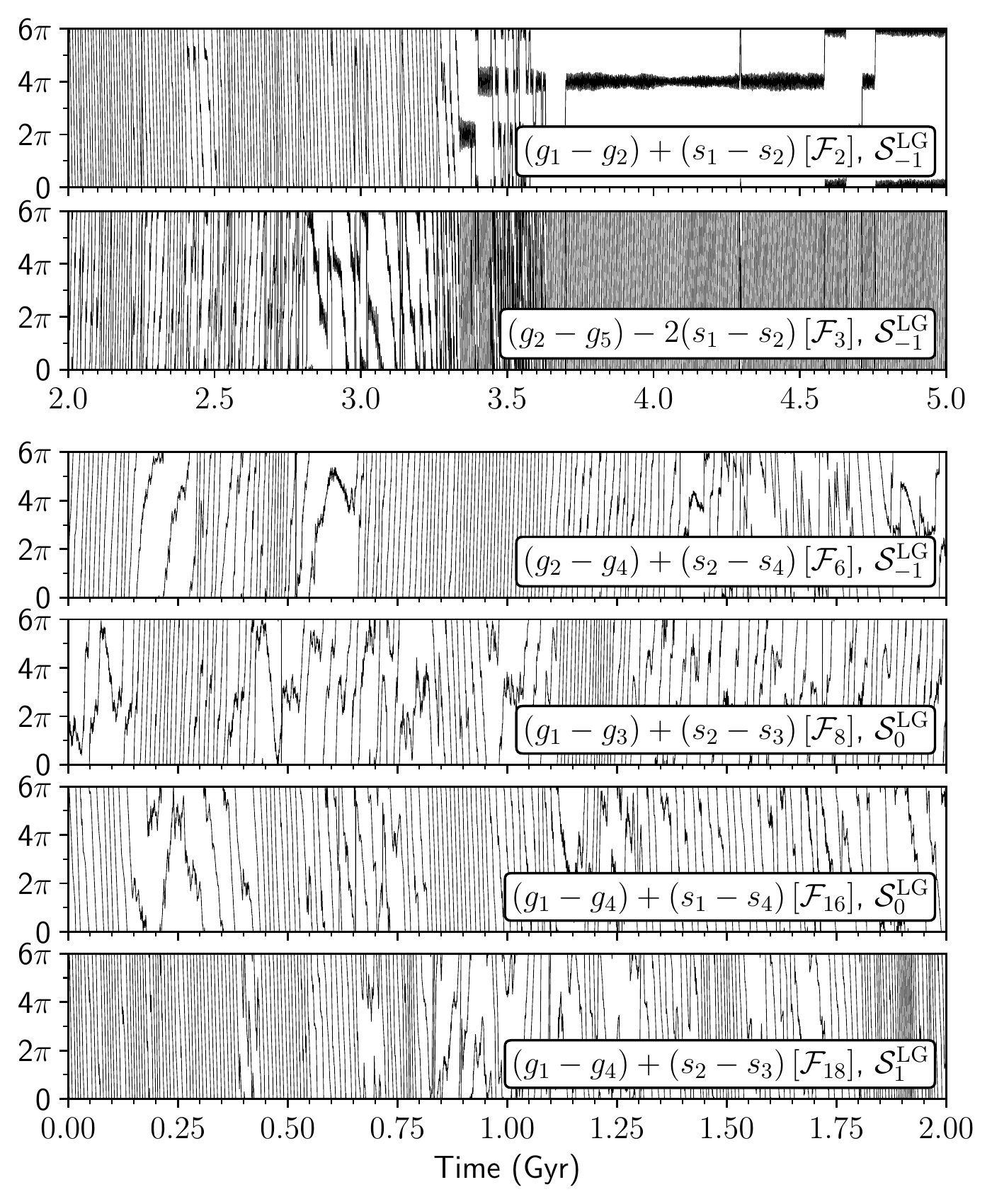}
\caption{Combination of angle variables for different harmonics of Table~\ref{tab:resonant_harmonics} 
as a function of time along direct integrations $\mathcal{S}^{\textrm{LG}}_n$ of the Solar System presented in \citet{Laskar2009} 
($n=0$ is the nominal solution). The angle variables correspond to the proper modes $\vec{z}^\bullet, \vec{\zeta}^\bullet$ 
defined in \citet{Laskar1990}. The time series are low-pass-filtered with a cutoff frequency of $(\textrm{500 kyr})^{-1}$.}
\label{fig:librations}
\end{figure}

The statistical nature of these findings suggests that the harmonics of Table~\ref{tab:resonant_harmonics} should be found 
statistically in resonance over an ensemble of sufficiently long orbital solutions of the Solar System, independently of 
the precise proper modes and dynamical model employed. 
A vivid example is provided by the term $(g_1-g_2)+(s_1-s_2)$ that we found among the top resonances: it enters libration 
at around 1 Gyr in \citet{Lithwick2011} but was not mentioned for the much shorter solutions of previous studies 
\citep{Laskar1990,Sussman1992,Laskar1992,Laskar2004}. 
More generally, we show in Fig.~\ref{fig:librations} examples of libration episodes for a number of harmonics in Table~\ref{tab:resonant_harmonics} 
along the direct integrations of the Solar System presented in \citet{Laskar2009} and spanning 5 Gyr, when one uses 
the proper modes $\vec{z}^\bullet, \vec{\zeta}^\bullet$ defined in \citet{Laskar1990}. Despite the different dynamical model and proper modes, 
one systematically confirms the librations predicted by the theory by inspecting just a few solutions. We point out the remarkable case of $(g_1-g_3)+(s_2-s_3)$, 
which will surely be librating over the next few tens of millions of years. Clearly, the chaotic zone of the ISS may extend to resonances 
that are barely, or not at all, active along our nominal solution and thus do not appear in Table~\ref{tab:resonant_harmonics}. 
In any case, Fig.~\ref{fig:librations} suggests that, despite being deduced from the forced secular dynamics, the web of resonances pictured 
here should underlie the chaotic dynamics of the inner planets in every realistic model of the Solar System. 

\begin{acknowledgements}
We are truly indebted to M.~Gastineau for his support with TRIP. We acknowledge 
the precious help of L.~Robol for the integration of MPSolve in TRIP, and the 
fruitful discussions with N.~H.~Hoàng and M.~Saillenfest. 
F.~M. has been supported by a PSL post-doctoral fellowship and by a grant of the 
French Agence Nationale de la Recherche (AstroMeso ANR-19-CE31-0002-01). 
This project has been supported by the European Research Council (ERC) 
under the European Union’s Horizon 2020 research and innovation program (Advanced Grant AstroGeo-885250). 
This work was granted access to the HPC resources of MesoPSL financed by the Region 
Île-de-France and the project Equip@Meso (reference ANR-10-EQPX-29-01) of the 
programme Investissements d’Avenir supervised by the Agence Nationale pour la Recherche. 
\end{acknowledgements}

%
\bibliographystyle{aa} 
\bibliography{iss2} 
%

\begin{appendix}

\section{The need for computer algebra. TRIP}
\label{supp:trip}
When it comes to elucidating the causes of a chaotic dynamics in celestial mechanics, 
one typically resorts to the analysis of the Fourier harmonics of a given Hamiltonian 
\citep{Chirikov1960,Chirikov1969,Walker1969,Wisdom1980}. 
The problem is first reduced to a set of partial one-DOF Hamiltonians that describe 
the dynamics arising from each harmonic independently of one another. The interaction of pairs of harmonics 
is then addressed through the resonance overlap criterion \citep{Chirikov1960,Walker1969,Chirikov1979}  
and Poincar\'{e}'s sections \citep{Henon1964} to reveal the dynamical sources of chaos. 
In practice, the choice of the harmonics to study may rely on a somewhat intuitive perception of 
their relevance, supported by observational data, numerical investigations, and the findings of 
previous research.

When the Hamiltonian system under investigation has a few DOFs, the above approach is 
quite straightforward to apply and very reliable \citep[e.g.][]{Wisdom1984}. Conversely, when dealing with several 
DOFs, as in case of the Solar System planets, the curse of dimensionality makes its implementation 
especially difficult. First of all, the number of low-order harmonics that are potentially resonant 
rapidly grows with the dimension of the phase space: an a priori choice of the pertinent harmonics may thus lead 
to a biased analysis. More importantly, the resonant nature of each harmonic, even in a one-DOF 
model, still depends on the system position in the full-dimensional phase space.
Some authors have invoked the quasi-periodicity of certain variables to reduce the analysis to 
a lower-dimension phase space \citep{Lithwick2011,Boue2012,Batygin2015}. However, the validity of such an 
assumption strongly depends on the problem under investigation, and it cannot provide the basis of a general 
approach. In the ISS, for instance, the fundamental frequencies of the planet orbits
all vary in an intricate way in a few tens of millions of years, with the exception of the eccentricity 
mode $g_2$, which has slower frequency variations \citep{Laskar1990,Laskar2004,Hoang2021}. A quasi-periodic approximation 
for the corresponding DOFs is thus not consistent over timescales much longer than 10 Myr \citep{Mogavero2021}. 

To overcome the difficulties associated with the high-dimensional phase space of the ISS, 
we set up an unbiased systematic analysis of the Fourier harmonics of the secular planetary Hamiltonian. 
To this end, we employed TRIP, a computer algebra system dedicated to the perturbation series of celestial mechanics, 
which has been developed over the past 30 years at the \emph{Institut de mécanique céleste et de calcul des éphémérides} 
\citep{Laskar1990b,Gastineau2011,TRIP}. The main objects of its symbolic kernel 
are the Poisson series, that is, multivariate Fourier series whose coefficients are multivariate Laurent series\footnote{Throughout 
the paper, $j = \sqrt{-1}$ stands for the imaginary unit and $\E$ represents the exponential operator.}, 
\begin{equation}
\label{eq:poisson_series}
S(z_1, \dots, z_n, \varphi_1, \dots, \varphi_m) = 
\sum C_{k,\ell} z_1^{k_1} \cdots z_n^{k_n} \E^{j \left(\ell_1 \varphi_1 + \dots + \ell_m \varphi_m \right)}
,\end{equation}
where $(z_p)_{p=1}^n$ and $(\varphi_p)_{p=1}^m$ are complex and real variables, respectively, 
$k = (k_p)_{p=1}^n \in \mathbb{Z}^n$ and $\ell = (\ell_p)_{p=1}^m \in \mathbb{Z}^m$. 
The angle variables $(\varphi_p)_{p=1}^m$ enter the series through their complex exponential, which is encoded in TRIP as a dedicated variable. 
The complex coefficients $C_{k,\ell}$ can be rational functions or numerical coefficients of different types: 
double-, quadruple-, and multiple-precision floating-point numbers; fixed or multiple-precision integers or rational numbers; 
or double- and quadruple-precision floating-point intervals. 
The symbolic kernel implements several operations on the Poisson series: sum, product, division, differentiation, integration, exponentiation, 
substitution of variables, and selection of specific terms. At the heart of these functionalities, TRIP embeds a monomial 
truncated product, which allows the product of two series  to be truncated according to a given condition on the partial 
or total degree of one or more variables (truncation based on the size of the coefficients $C_{k,\ell}$ is also 
available). Truncations are formal objects in TRIP and permit perturbative computations to be performed at a 
minimal cost. TRIP also provides special functionalities for celestial mechanics, such as 
Poisson brackets and the manipulation of formal Lie series, which allow the algorithms of 
canonical perturbation theory \citep{Hori1966,Deprit1969} to be implemented. A numerical kernel, based on vectors, matrices, 
and multidimensional tables, provides a state-of-the-art evaluation of the Poisson series. It also includes 
a number of routines typical of more general computer algebra systems, such as algorithms for the zeros 
of univariate polynomials, the solutions of systems of algebraic equations, and the analysis of time 
series (e.g. fast Fourier transform, frequency analysis, integration, and interpolation). Moreover, 
TRIP natively integrates polynomial dynamical systems through Adams PECE and DOPRI8 methods \citep{HairerI,Prince1981}. 
Finally, TRIP provides a dedicated procedural language that supports loops, conditional statements, 
and function and structure definition, which permits the symbolic and numerical kernels to be interfaced in autonomous programs. 
When unsupported functionalities are needed, it allows external C and Fortran routines to be called and allows communication 
with other computer algebra systems, such as Maple and Mathematica. 

\section{Development of the secular planetary Hamiltonian}
\label{supp:model}
The secular Hamiltonian $\Hsec$ of the entire Solar System in Eq.~\eqref{eq:ham_sec} can be expanded in series of the \Poincare{} complex 
variables and truncated at a given total degree $2n$, where $n$ is a positive integer \citep{Laskar1991,Laskar1995}. 
This results in a polynomial Hamiltonian $\Hsec_{2n} = \sum_{p = 0}^n \Hsec_{(2p)}$, 
where $\Hsec_{(2p)}$ groups all the monomials of same total degree $2p$ in the variables 
$(x_k, \bar{x}_k, y_k, \bar{y}_k)_{k=1}^8$. The expansion straightforwardly provides the 
truncated Hamiltonian for the forced ISS, 
\begin{equation}
\label{eq:ham_iss_expansion}
\begin{aligned}
&\Hiss_{2n} =
\sum_{p = 1}^n \Hiss_{(2p)}, \\
&\Hiss_{(2p)}[(x_k,y_k)_{k=1}^4,t] = 
\Hsec_{(2p)}[(x_k,y_k)_{k=1}^4, (x_k = x_k(t), y_k = y_k(t))_{k=5}^8]. 
\end{aligned}
\end{equation}
At the lowest degree, the $\Hiss_2$ Hamiltonian describes an integrable, forced LL dynamics for 
the inner planets. The corresponding equations of motion are analytically solved by introducing complex 
variables $(u_k, v_k)_{k=1}^4$, with $(u_k, -j \bar{u}_k; $ $v_k, -j \bar{v}_k)$ conjugate 
momentum-coordinate pairs, that correspond to the proper (i.e. normal) modes of the free oscillations \citep{Mogavero2021}. 
By switching to the new set of canonical variables $(u_k, v_k)$, the truncated Hamiltonian 
in Eq.~\eqref{eq:ham_iss_expansion} transforms to 
\begin{equation}
\label{eq:ham_sec_inn_uv}
\Huv_{2n}[(u_k, v_k),t] = 
\sum_{p = 1}^{n} \Huv_{(2p)}[(u_k, v_k),t] 
,\end{equation}
where $\Huv_{(2p)}$ groups all the terms of same total degree $2p$ \citep{Mogavero2021}. 
Action-angle variables $(\Chi_k,\chi_k;$ $\Psi_k,\psi_k)_{k=1}^4$ that trivially integrate the LL problem 
are thus introduced as 
\begin{equation}
\label{eq:AA_vars}
u_k = \sqrt{\Chi_k} \, \E^{-j \chi_k}, \quad
v_k = \sqrt{\Psi_k} \, \E^{-j \psi_k} 
\end{equation}
for $k \in \{1,2,3,4\}$ and allow the truncated Hamiltonian in Eq.~\eqref{eq:ham_sec_inn_uv} to be written as a finite Fourier series, 
\begin{equation}
\label{eq:fourier_expansion}
\Huv_{2n}(\vecI,\vectheta,t) = 
\sum_{\vec{k} \in \mathbb{Z}^8}
\sum_{\vec{\ell} \in \mathbb{Z}^7} 
\fourcoeff{2n}{k}{\ell} (\vecI)
\E^{j \left( \vec{k} \cdot \vec{\theta} + \vec{\ell} \cdot \out{\vec{\omega}} t \right)} 
,\end{equation}
where $\widetilde{\Huv}_{2n}^{\vec{k},\vec{\ell}}$ are complex amplitudes, with $\widetilde{\Huv}_{2n}^{-\vec{k},-\vec{\ell}} = 
\overline{\fourcoeff{2n}{k}{\ell}}$, and where we employ a compact notation for the action-angle variables, that is, 
$\vecI = (\vecChi,\vecPsi)$ and $\vectheta = (\vecchi,\vecpsi)$.
At the lowest degree, one has the LL Hamiltonian $\Huv_2 = \fourcoeff{2}{0}{0} = \vecomegaLL \cdot \vecI$, 
with $\vecomegaLL = -(\vec{g}_{\text{LL}}, \vec{s}_{\text{LL}})$, and the Hamiltonian in 
Eq.~\eqref{eq:fourier_expansion} is therefore in quasi-integrable form. 
In the framework of canonical perturbation theory, the explicit time dependence of the Hamiltonian 
(coming from the forcing of the outer planets) can be absorbed in a phase-space extension, resulting in 
\begin{equation}
\label{eq:extended_phasespace}
\Huv^\star_{2n}(\vecI,\vectheta;\vec{\Phi},\vec{\phi}) = 
\out{\vec{\omega}} \cdot \vec{\Phi} 
+ \sum_{\vec{k} \in \mathbb{Z}^8}
\sum_{\vec{\ell} \in \mathbb{Z}^7} 
\fourcoeff{2n}{k}{\ell}(\vecI) 
\E^{j \left( \vec{k} \cdot \vec{\theta} + \vec{\ell} \cdot \vec{\phi} \right)} 
.\end{equation}
The dynamics of the additional angles is consistently given by $\dot{\vec{\phi}} = \out{\vec{\omega}}$, 
while that of the conjugated actions $\vec{\Phi}$ is irrelevant. 

\section{Nominal solution and choice of initial conditions}
\label{supp:nomsol}
We estimated the statistical properties of the chaotic dynamics in the ISS from the orbital solution \nomsol{} 
of Gauss's dynamics (Hamiltonian $\Hiss$ in Eq.~\eqref{eq:ham_sec_inn}) that is numerically integrated 
over 5 Gyr starting from the nominal initial conditions given in \citet[Appendix D]{Mogavero2021}. 
This solution is denoted as the nominal solution throughout the paper and consists of the values of the canonical variables 
sampled with a timestep $\Delta t$ of 1000 yr, that is, 
\nomsol{} $= \{\mathcal{S}(t_m)\}_{m=0}^{5 \cdot 10^6}$ with \nomsol{}$(t_m) = (\vecI(t_m),\vectheta(t_m))$ and 
$t_m=m\Delta t$. 
We considered the time distribution of a physical observable along the nominal solution \nomsol{} 
as representative of its ensemble distribution (that is, over a set of stable orbital solutions with very close initial conditions) 
at some large time of the order of billions of years. 

Following \citet{Mogavero2021}, we chose the initial conditions for the ensemble computation of the FT-MLE 
by taking the relative variation in each coordinate of the nominal phase-space vector of Gauss's dynamics
as a normal random variable, with zero mean and a standard deviation of $10^{-9}$. In other words, the initial 
conditions are distributed according to
\begin{equation}
\label{eq:IC}
x_i = x_i^\ast + \sigma \left( \operatorname{Re}\{x_i^\ast\} \, z_i + j \operatorname{Im}\{x_i^\ast\} \, z_i^\prime \right) 
\end{equation}
for $k \in \{1,2,3,4\}$, where $x_i^\ast$ represents the initial conditions of the nominal solution \nomsol{}, 
$z_i, z_i^\prime \sim \mathcal{N}(0,1)$ are standard normal deviates, and $\sigma = 10^{-9}$.
An analogous expression holds for the variables $y_i$. The initial conditions thus follow a multivariate 
Gaussian distribution centred at the nominal initial conditions of $\Hiss$, with a diagonal covariance matrix. 
The initial tangent vectors that we used in the application of the \citet{Benettin1980} algorithm are also sampled 
from a multivariate Gaussian distribution, with null mean and an identity covariance matrix. 
We chose a renormalisation time of 5 Myr and the Euclidean norm for the phase-space vectors. 

\section{Leading harmonics}
\label{supp:leading_harm}
There can be many different ways of ranking the Fourier harmonics of Eq.~\eqref{eq:fourier_expansion}, depending on the dynamical quantities that one aims to track. As we intend to reconstruct the resonant structure of the ISS, we are interested in the long-term variation in the frequencies of motion $\vec{\omega} = \dot{\vectheta}$. From Eq.~\eqref{eq:fourier_expansion}, Hamilton's equations give 
\begin{equation}
\label{eq:omega_freqs}
\vec{\omega}_{2n} = 
\frac{\partial \fourcoeff{2n}{0}{0}}{\partial \vecI} 
+ \sideset{}{^\star}\sum_{\vec{k},\vecl}
\frac{\partial \fourcoeff{2n}{k}{\ell}}{\partial \vecI}
\E^{j \left( \vec{k} \cdot \vec{\theta} + \vecl \cdot \out{\vec{\omega}} t \right)}
,\end{equation}
where $(\vec{k},\vecl) \in \mathbb{Z}^8 \times \mathbb{Z}^7$ and the star means that the null wave vector $(\vec{0},\vec{0})$ is excluded from the summation. As the Hamiltonian $\Huv_{2n}$ is close to integrable, the actions $\vecI(t)$ vary over a timescale much longer than that of the angles $\vectheta(t)$. Therefore, Eq.~\eqref{eq:omega_freqs} suggests that the frequencies $\vec{\omega}_{2n}(t)$ rapidly fluctuate around the slow-varying mean frequencies $\vecomegamean_{2n}(\vecI (t)) = \partial \fourcoeff{2n}{0}{0} / \partial \vecI$, only depending on the action variables. If the motion were quasi-periodic, the harmonics appearing in the summation of Eq.~\eqref{eq:omega_freqs} could be systematically eliminated by a series of canonical transformations in the form of Lie transforms. The mean frequencies would then give the (constant) frequencies of motion as a function of the transformed action variables. For a chaotic dynamics, the presence of resonant harmonics shifts the short-time average of $\vec{\omega}_{2n}$ with respect to $\vecomegamean_{2n}$. Nevertheless, the mean frequencies still provide the overall long-term behaviour of the frequencies of motion. We thus chose to rank the Fourier harmonics of the Hamiltonian according to their contribution to the system trajectory in the action space, which in turn drives the variation in the mean frequencies with time. 

Hamilton's equation $\vec{\dot{I}} = - \partial \Huv_{2n} / \partial \vectheta$ gives the contribution of the harmonic $\vec{k},\vec{\ell}$ to the action vector $\vecI(t)$ as 
\begin{equation}
\Delta \vecI_{\vec{k},\vecl}(t) = 
2 \vec{k} \operatorname{Im} 
\int_0^t 
dt^\prime \ \fourcoeff{2n}{k}{l}(\vecI(t^\prime)) \E^{j \left( \vec{k} \cdot \vec{\theta}(t^\prime) + \vecl \cdot \out{\vec{\omega}} t^\prime \right)} 
.\end{equation}
By construction, the sum of the contributions from all the harmonics exactly reconstructs the system displacement in the action space at a given time, that is, $\Delta \vecI(t) = \vecI(t) - \vecI(0)$. The integral was computed numerically in TRIP for each harmonic along the sampled nominal solution \nomsol{}. The harmonic ranking was established by considering the relative Euclidean norm of the contribution, 
\begin{equation}
\label{eq:harmonic_ranking_2}
\delta\vecI_{\vec{k},\vecl}(t) = 
\frac{\left\Vert \Delta \vecI_{\vec{k},\vecl}(t) \right\Vert}{\left\Vert \vecI(t) \right\Vert}
.\end{equation}
The set of values $\delta\vecI_{\vec{k},\vecl} = \{\delta\vecI_{\vec{k},\vecl}(t_m)\}_{m=0}^{5 \cdot 10^6}$ is treated as a sample 
from an underlying PDF and characterised via statistical estimators. The harmonics in Table~\ref{tab:leading_harmonics} 
are listed according to their median contribution, while the 5$^{\mathrm{th}}$ and 95$^{\mathrm{th}}$ percentiles show 
the dispersion of $\delta\vecI_{\vec{k},\vecl}$. 

\section{Partial Hamiltonians}
\label{supp:partial_harm}
\begin{figure}
\includegraphics[width=\columnwidth]{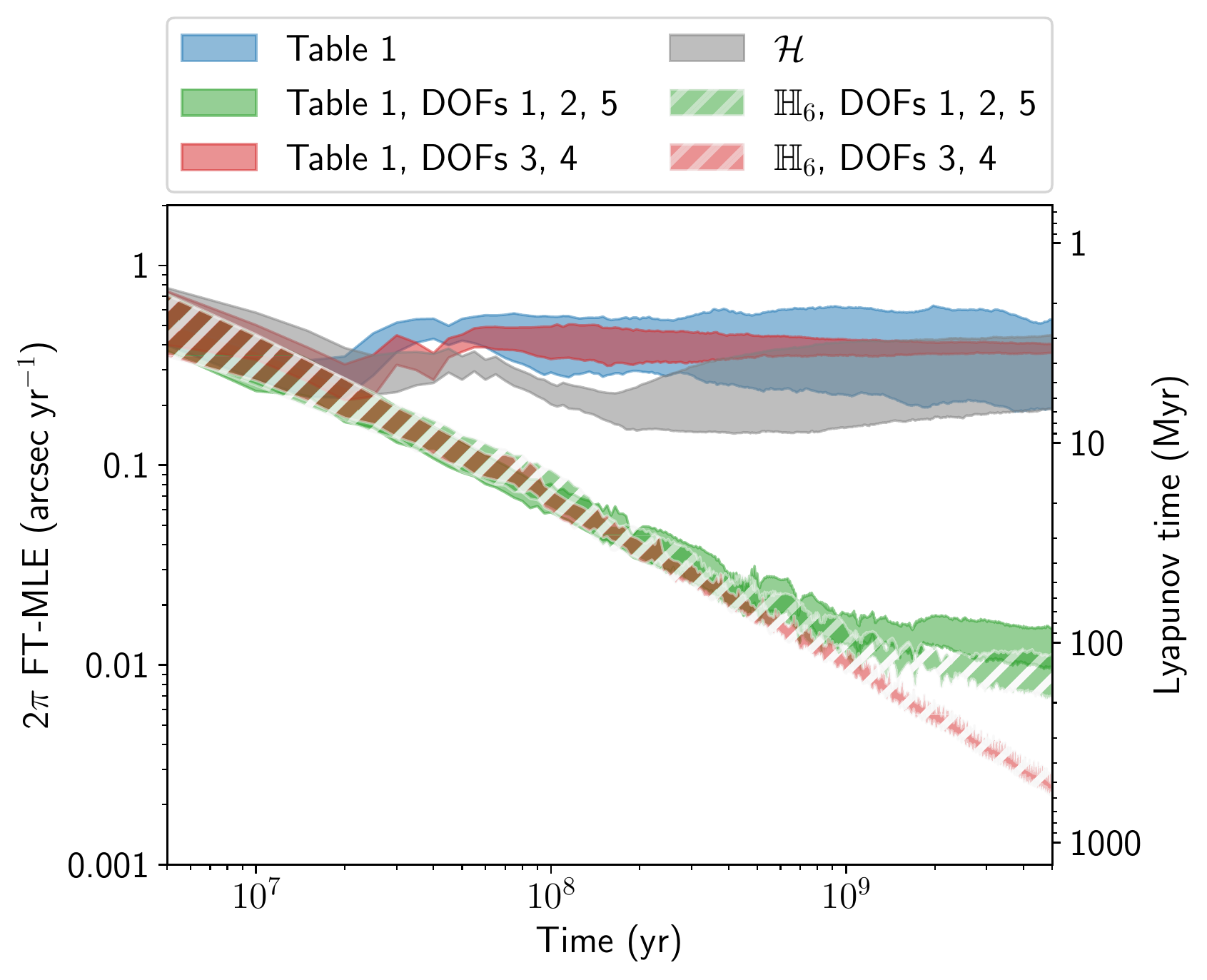}
\caption{Finite-time MLE and corresponding Lyapunov time FT-MLE$^{-1}$ from different 
partial Hamiltonians (Eq.~\eqref{eq:partial_ham}). 
The bands represent the [5$^{\textrm{th}}$, 95$^{\textrm{th}}$] percentile range of the PDF estimated 
from an ensemble of 128 stable orbital solutions with very close initial conditions.  
The blue colour corresponds to the Fourier harmonics of Table~\ref{tab:leading_harmonics}, 
the green colour to the harmonics of the same table that only include the angular DOFs 1, 2, and 5, 
and the red colour to those harmonics that only involve the angular DOFs 3 and 4. 
The hatched regions correspond to the case where the choice of the DOFs is made across the entire $\Huv_6$ Hamiltonian. 
The grey region stands for the full $\Hiss$ Hamiltonian, as shown in Fig.~\ref{fig:lyap}.}
\label{fig:lyap_partials}
\end{figure}

The identification of the leading harmonics of the Hamiltonian allows the FT-MLE to be reproduced
from a small set $\mathcal{U}$ of Fourier harmonics, by considering 
\begin{equation}
\label{eq:partial_ham}
\Huv_{6}^\mathcal{U} = 
\fourcoeff{6}{0}{0}
+ \sideset{}{^\star}\sum_{(\vec{k},\vecl) \in \mathcal{U}}
\fourcoeff{6}{k}{\ell} 
\E^{j \left( \vec{k} \cdot \vec{\theta} + \vecl \cdot \out{\vec{\omega}} t \right)}
.\end{equation}
Partial Hamiltonians $\Huv_{6}^\mathcal{U}$ can be systematically constructed in TRIP through the selection of 
specific terms of $\Huv_6$. We integrated the corresponding variational equations to obtain the FT-MLE 
distribution of an ensemble of 128 stable solutions with initial conditions very close to the nominal ones of Gauss's dynamics. 
Figure \ref{fig:lyap_partials} shows the [5$^{\textrm{th}}$, 95$^{\textrm{th}}$] percentile range of the FT-MLE distribution 
from the 30 leading harmonics of Table~\ref{tab:leading_harmonics}. This simplified dynamics reproduces the asymptotic FT-MLE distribution of the Hamiltonians $\Hiss$ and $\Hiss_{2n}$  
very well, confirming the dynamical relevance 
of the leading harmonics. Very importantly, the FT-MLE prediction turns out to be robust with respect to the addition of 
further harmonics from our ranking. 

The construction of partial Hamiltonians allows the extensive entanglement of the DOFs already shown 
in Table~\ref{tab:leading_harmonics} to be highlighted. Figure \ref{fig:lyap_partials} reports the FT-MLE distribution 
from the leading harmonics of Table~\ref{tab:leading_harmonics} that only involve the angles of the proper modes 1, 2, and 5, 
that is, the harmonics $\mathcal{U}~=~\{\F_6, \F_8, \F_{20}, \F_{27}, \F_{29}, \F_{30}\}$ only involving combinations of 
$g_1,g_2,g_5,s_1,$ and $s_2$. This partial Hamiltonian includes, in particular, the Fourier harmonics considered in the 
simplified long-term dynamics of Mercury in \citet{Batygin2015}. The predicted Lyapunov time of about 100 Myr 
is clearly incompatible with the full system dynamics. The prediction does not improve when the selection of the harmonics 
only involving $g_1,g_2,g_5,s_1,$ and $s_2$ is made across the entire $\Huv_6$ Hamiltonian (as shown by the hatched green region). 
The harmonics of Table~\ref{tab:leading_harmonics} that only involve the angles of the proper modes 3 and 4, that is, 
the harmonics $\mathcal{U}=\{\F_2, \F_3, \F_5, \F_{25}\}$ only involving $g_3,g_4,s_3$ and $s_4$, produce an 
asymptotic FT-MLE that is compatible with the full system dynamics (as shown  by the red region in Fig.~\ref{fig:lyap_partials}). 
Nevertheless, when the selection of such harmonics is made across the entire $\Huv_6$ Hamiltonian, the resulting 
FT-MLE turns out to be a monotonically decreasing function of time, suggesting an integrable dynamics (hatched red region 
in Fig.~\ref{fig:lyap_partials}). These numerical experiments show that freezing a priori some DOFs may dramatically 
change the behaviour of the system, resulting in predictions that may largely depend on the specific choice of the Fourier harmonics. 
Clearly, the surprising behaviours of Fig.~\ref{fig:lyap_partials} may depend on the fact that the initial conditions 
of the simplified dynamics are not adjusted according to the choice of the harmonics. Proper initial conditions should be 
computed from the Lie transform that allows the harmonics of the Hamiltonian $\Huv_{6}$ that do not appear 
in $\Huv_{6}^\mathcal{U}$  to be eliminated. However, this adjustment is often omitted when simplified models are constructed in literature. 
Due to this fact, we decided to keep the nominal initial conditions of Gauss's dynamics in these numerical experiments. 
In any case, taking the ensemble of the DOFs of the inner system into account seems essential. The resonant nature 
of a Fourier harmonic, in particular, should only be established along the orbital solution of a realistic Hamiltonian. 

\section{Lie transform}
\label{supp:lie}
\label{supp:librating_harm}
We employed canonical perturbation theory to construct a change of variables that eliminates the Fourier harmonics of degree 4 
from the truncated Hamiltonian, that is, the harmonics with a non-null wave vector appearing in $\Huv_{(4)}$ in Eq.~\eqref{eq:ham_sec_inn_uv}. 
Such a transformation of variables can be canonically defined as the time-1 flow of a generating Hamiltonian $\mathrm{S}$ 
that satisfies the homologic equation 
\begin{equation}
\label{eq:homol_eq}
\Ham{(4)} + \{\mathrm{S}, \Ham{2} + \out{\vec{\omega}} \cdot \vec{\Phi}\} = \fourcoeff{(4)}{0}{0} 
,\end{equation}
where the braces represent the Poisson bracket and where we employ the extended phase-space formalism of Eq.~\eqref{eq:extended_phasespace} 
to deal with a time-independent Hamiltonian. The resulting change of variables is given by the formal Lie transform 
\begin{equation}
\label{eq:hat_vars}
(\lieI, \lietheta; \liePhi, \liephi) = \E^{-L_\mathrm{S}} (\vecI, \vectheta; \vec{\Phi}, \vec{\phi})
,\end{equation} 
where $L_\mathrm{S} \, \cdot = \{\mathrm{S}, \cdot\}$ is the Lie derivative associated with the generating function $\mathrm{S}$. 
The exponential of an operator $A$ acting on the phase-space functions is formally expressed as 
\begin{equation}
\label{eq:operator_E}
\E^A = \sum_{q=0}^{+\infty} \frac{A^q}{q!}  
,\end{equation}
with $A^0$ defined as the identity operator and $A^q = A A^{q-1}$ for $q \geq 1$. 
The transformed Hamiltonian reads 
\begin{equation}
\label{eq:lie_transf}
\lieham(\lieI, \lietheta; \liePhi, \liephi) = 
\E^{L_\mathrm{S}} \, \Hamstar{2n} \Big|_{\lieI, \lietheta; \liePhi, \liephi} 
,\end{equation}
where the Lie transform is formally evaluated at the new canonical variables $\lieI, \lietheta; \liePhi, \liephi$. 
The solution to the homologic Eq.~\eqref{eq:homol_eq} is expressed as 
\begin{equation}
\label{eq:gen_func}
\mathrm{S}(\vecI, \vectheta; \vec{\phi}) = - j 
\sideset{}{^\star}\sum_{\vec{k},\vecl}
\frac{\fourcoeff{(4)}{k}{l}(\vecI)}{\vec{k} \cdot \vecomegaLL + \vecl \cdot \out{\vec{\omega}}} 
\E^{j \left( \vec{k} \cdot \vec{\theta} + \vec{\ell} \cdot \vec{\phi} \right)} 
.\end{equation}
As $\Ham{(4)}$ possesses a finite number of non-null harmonics, the Fourier series in Eq. \eqref{eq:gen_func} is finite. Moreover, 
all the denominators being constant and different from zero, the generating function $\mathrm{S}$ is a well-defined analytical function. 

Because the generating function $\mathrm{S}$ in Eq.~\eqref{eq:gen_func} does not depend on the actions $\vec{\Phi}$, 
the only term appearing in $\lieham$ that depends on the transformed variables $\liePhi$ is $\out{\vec{\omega}} \cdot \liePhi$. 
The values of $\liePhi$ are therefore irrelevant from a dynamical point of view (as are those of $\vec{\Phi}$), 
and the transformation $\liePhi = \E^{-L_\mathrm{S}} \vec{\Phi}$ can be discarded. Moreover, from $\{\mathrm{S}, \vec{\phi}\} = \vec{0}$ 
it follows that $\liephi = \vec{\phi}$, in agreement with the previous point. 

To implement the exponential operator in TRIP, the formally infinite summation in Eq. \eqref{eq:operator_E} must be truncated. 
As all the terms in the generating function $\mathrm{S}$ are of degree 4, the Lie derivative $L_\mathrm{S}$ raises the degree 
of the Hamiltonian terms by 2. Therefore, to truncate the transformed Hamiltonian at degree $2n$, it is sufficient 
to compute the summation up to the order $n-1$ and then discard the terms of degree higher than $2n$, that is, 
\begin{equation}
\label{eq:lie_transf_trunc}
\lieham_{2n} \xleftarrow{\textrm{Truncation at degree } 2n} \left( \sum_{q=0}^{n-1} \frac{L_\mathrm{S}^q}{q!} \right) \Hamstar{2n} 
.\end{equation}
Once the Hamiltonian $\lieham_{2n}$ is computed, the corresponding nominal solution $\mathcal{S}^\prime_{2n}$ is obtained 
from the truncation of the operator $\E^{-L_\mathrm{S}}$ at the same order $n-1$, that is, 
$\mathcal{S}^\prime_{2n} = \left( \sum_{q=0}^{n-1} (-1)^q L_\mathrm{S}^q / q! \right) \mathcal{S}$. 
Computing the change in the nominal solution is essential for consistently evaluating the amplitudes of the Hamiltonian terms 
resulting from the Lie transform. 

\begin{figure}
\includegraphics[width=\columnwidth]{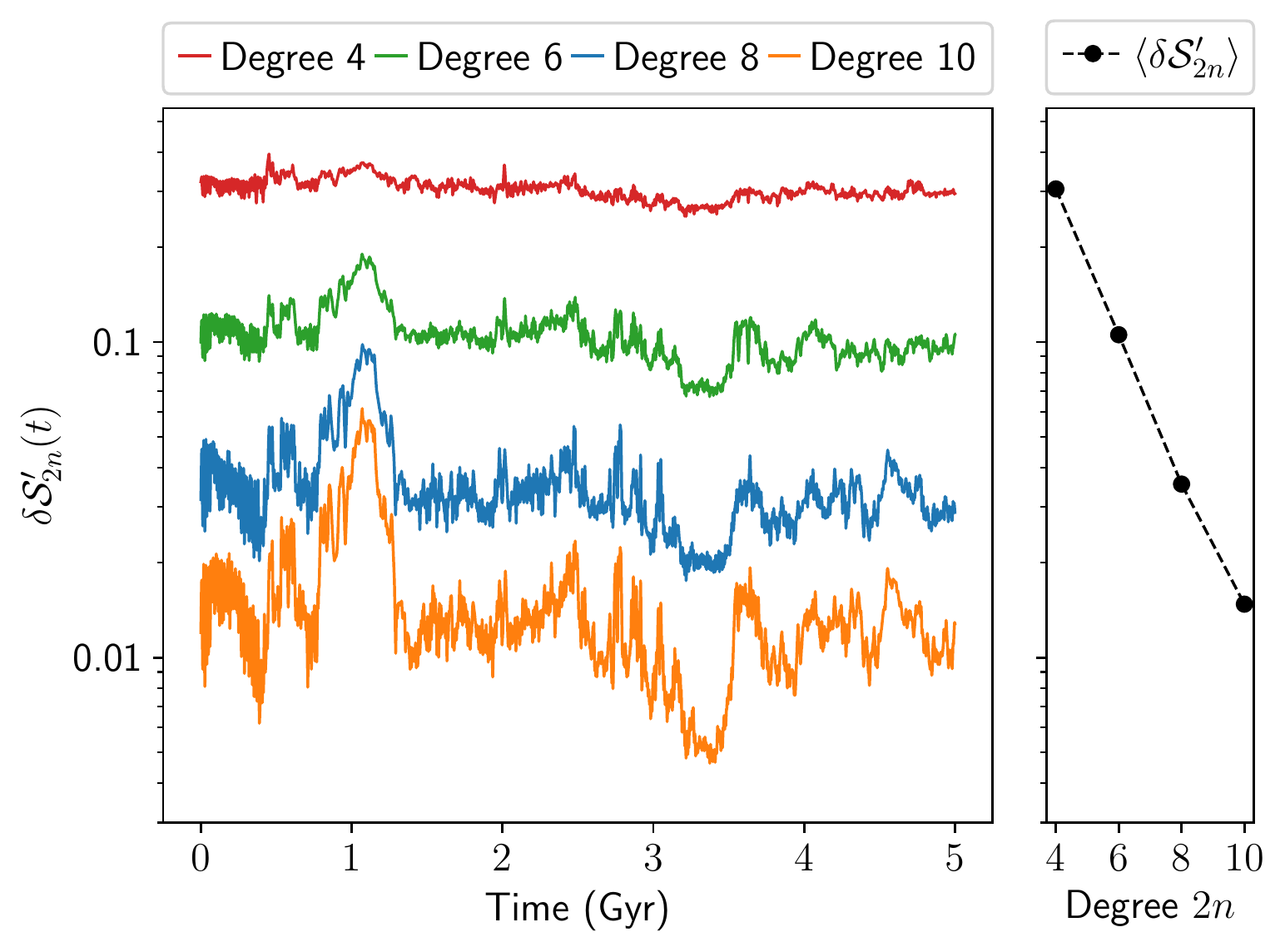}
\caption{Convergence of the Lie-transformed nominal solution $\mathcal{S}^\prime_{2n}$. 
Left panel: Relative increments $\delta \mathcal{S}^\prime_{2n}(t) = 
\left\Vert \mathcal{S}^\prime_{2n}(t) - \mathcal{S}^\prime_{2n-2}(t) \right\Vert \big / \left\Vert \mathcal{S}^\prime_{2n}(t) \right\Vert$ 
as a function of time and truncation degree $2n$. Right panel: Corresponding time means $\langle \delta \mathcal{S}^\prime_{2n}(t) \rangle$. 
The time series are low-pass-filtered with a cutoff frequency of (5 Myr)$^{-1}$.}
\label{fig:lie_solution}
\end{figure}

\begin{figure}
\includegraphics[width=\columnwidth]{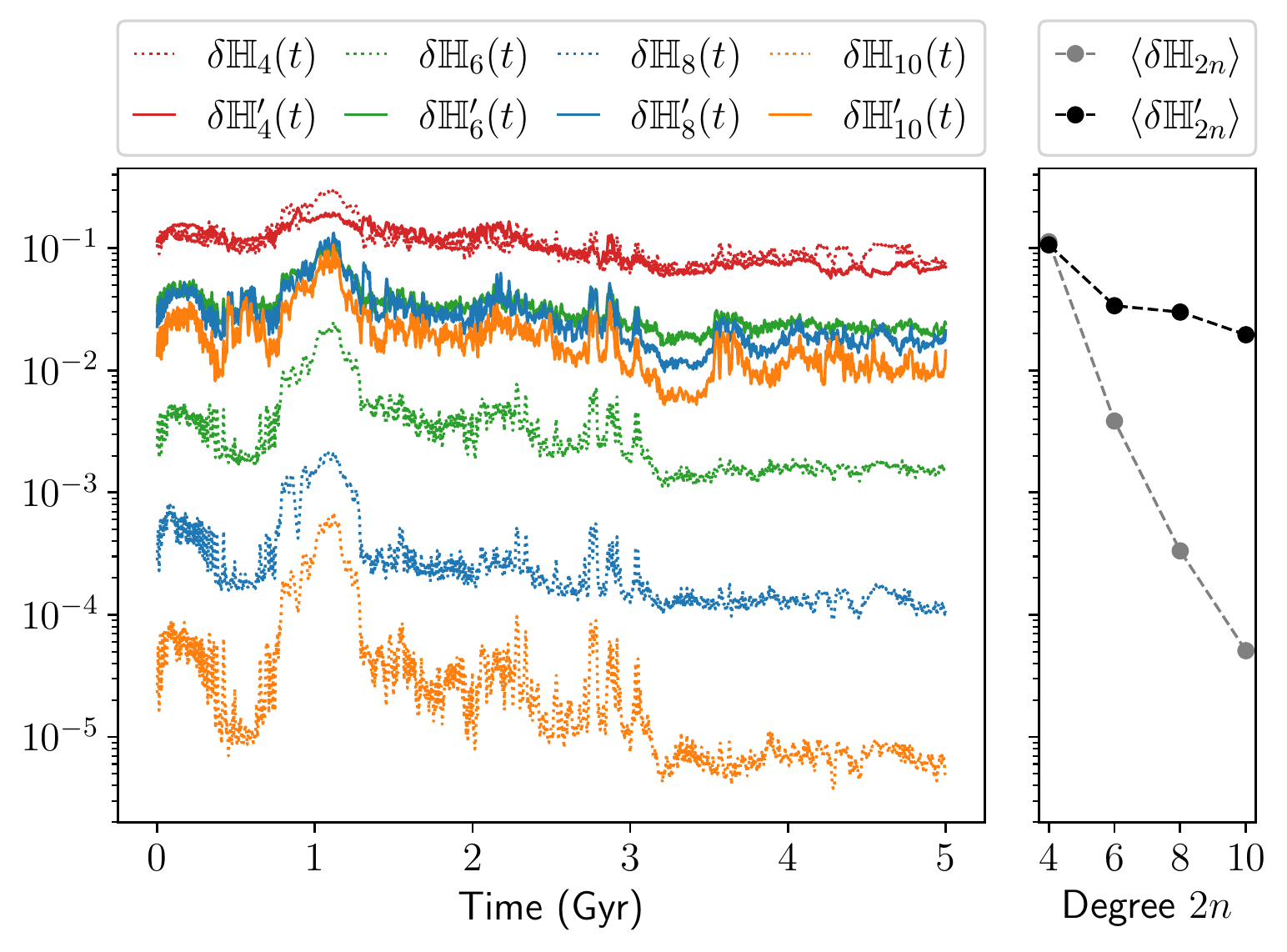}
\caption{Convergence of the Lie-transformed Hamiltonian $\lieham_{2n}$ along the nominal solution $\mathcal{S}^\prime_{10}$. 
Left panel: Relative increments $\delta \lieham_{2n}(t) = 
\left| \lieham_{2n}(\mathcal{S}^\prime_{10}(t)) - \lieham_{2n-2}(\mathcal{S}^\prime_{10}(t)) \right| \, \big / \, 
\big \langle \left| \lieham_{2n}(\mathcal{S}^\prime_{10}(t)) \right| \big \rangle$ 
as a function of time and truncation degree $2n$.  
Right panel: Corresponding time means $\langle \delta \lieham_{2n}(t) \rangle$. 
The time series are low-pass-filtered with a cutoff frequency of (5 Myr)$^{-1}$. 
The convergence of the original $\Huv_{2n}$ Hamiltonian along the nominal solution \nomsol{} is also shown.}
\label{fig:lie_hamiltonian}
\end{figure}

\subsection{Convergence}
Since the generating Hamiltonian $\mathrm{S}$ is an analytical function, the convergence of the series generated by the Lie 
transforms $\E^{\pm L_\mathrm{S}}$ depends on the size of its Fourier coefficients \citep{Morbidelli2002}. 
To address the convergence of our Lie transform, we first plotted the relative increments 
$\delta \mathcal{S}^\prime_{2n}(t) = 
\left\Vert \mathcal{S}^\prime_{2n}(t) - \mathcal{S}^\prime_{2n-2}(t) \right\Vert / \left\Vert \mathcal{S}^\prime_{2n}(t) \right\Vert$ 
of the transformed nominal solution as a function of time and truncation degree (see Fig.~\ref{fig:lie_solution}). We used the Euclidean norm for the phase-space vectors, as usual. 
To highlight the long-term trend of the time series, which may be hidden by large short-time oscillations, we applied the low-pass 
Kolmogorov-Zurbenko (KZ) filter with three iterations of the moving average and a cutoff frequency of (5 Myr)$^{-1}$ \citep{Zurbenko2017,Mogavero2021}. 
As also shown by the time mean of the increments $\langle \delta \mathcal{S}^\prime_{2n}(t) \rangle$, the contributions 
to $\mathcal{S}^\prime_{2n}$ decay exponentially with the degree of truncation, up to degree ten at least. That said, the decrease is 
in practice quite slow. Figure \ref{fig:lie_hamiltonian} then addresses the convergence of the Lie-transformed Hamiltonian $\lieham_{2n}$ 
along the transformed nominal solution $\mathcal{S}^\prime_{10}$. The relative increments $\delta \lieham_{2n}(t) = 
\left| \lieham_{2n}(\mathcal{S}^\prime_{10}(t)) - \lieham_{2n-2}(\mathcal{S}^\prime_{10}(t)) \right| \, \big / \, 
\big \langle \left| \lieham_{2n}(\mathcal{S}^\prime_{10}(t)) \right| \big \rangle$ are plotted as a function of time and truncation degree $2n$. 
We also report the convergence of the original $\Huv_{2n}$ Hamiltonian along the nominal solution \nomsol{} for comparison. 
Again, the contributions to $\lieham_{2n}$ decrease with the degree of truncation, up to degree ten at least, even though the decay is slow 
when compared to that of the original $\Huv_{2n}$ Hamiltonian. 
Even if the asymptotic convergence cannot be guaranteed, the absence of any divergence in Figs.~\ref{fig:lie_solution} and \ref{fig:lie_hamiltonian} 
suggests the stability of the estimations based on the truncation of the Lie transform at a finite degree $2n \leq 10$. Some oscillations 
of these estimations can be expected as a result of the slow (finite-degree) convergence of the series. 

\subsection{Librating harmonics}
To retrieve the set of librating harmonics, we defined time-dependent fundamental frequencies $(g_i(t_m), s_i(t_m))_{i=1}^4$ along 
a sampled orbital solution, as we did in \citet{Mogavero2021} in the case of the frequency $g_1$. 
Following Eq.~\eqref{eq:AA_vars}, the instantaneous (angular) frequencies of the proper modes $(u_i(t), v_i(t))$ are given by 
$(-\dot{\chi}_i(t), -\dot{\psi}_i(t))$. We sampled the time derivatives via the Lagrange five-point formula, as very high numerical 
precision is unnecessary. Since the instantaneous frequencies present large short-time fluctuations that hide their long-term 
evolution, we applied the low-pass KZ filter \citep{Zurbenko2017,Mogavero2021} to the time series and defined 
$(g_i(t_m), s_i(t_m)) = \textrm{KZ}[(-\dot{\chi}_i(t_m), -\dot{\psi}_i(t_m))]$. We used three iterations of the moving average 
and a cutoff frequency of (5 Myr)$^{-1}$. This specific value was chosen to filter out the non-chaotic part of the dynamical spectrum 
from the time series. When the fundamental frequencies were computed for a given orbital solution, the librating harmonics were defined 
from the combinations of frequencies that become null at least once over its time span. We find around 1000, 15\,000, and 150\,000 
librating harmonics among those appearing in $\lieham_{2n}$ along the Lie-transformed nominal solution $\mathcal{S}^\prime_{2n}$ at degree six, eight, and ten, respectively. 

\section{Reduced Hamiltonians}
\label{supp:reduced_ham}
To retrieve the resonant harmonics of the Lie-transformed Hamiltonian $\lieham_{2n}$, that is, 
\begin{equation}
\lieham_{2n}(\lieI, \lietheta; \liePhi, \liephi) = 
\out{\vec{\omega}} \cdot \liePhi 
+ \sum_{\vec{k},\vecl}
\liefourcoeff{k}{\ell} (\lieI) 
\E^{j(\vec{k} \cdot \lietheta + \vec{\ell} \cdot \liephi)} 
,\end{equation}
we considered the dynamics generated by each Fourier component separately. 
If we assume that at time $t=t_0$ along the nominal solution $\mathcal{S}^\prime_{2n}(t)$ the system is at position 
$(\lieI_0, \lietheta_0)$ in the phase space, the dynamics that would be generated by the harmonic $\veck,\vecl$, 
starting at $(\lieI_0, \lietheta_0, t_0)$, if all the other harmonics are ignored, is given by the reduced Hamiltonian 
\begin{equation}
\label{eq:red_ham}
h_{2n}^{\veck,\vecl}(\lieI,\lietheta,\tau) = \liefourcoeff{0}{0}(\lieI) 
+  2 \operatorname{Re} \{ \liefourcoeff{k}{\ell}(\lieI) \, \E^{j(\veck \cdot \lietheta + \Omega \tau)} \}
,\end{equation}
where $\tau \geq t_0$ is the time of the fictitious dynamics, $\Omega = \vecl \cdot \vecomegaout$, and 
we consider $\veck \in \mathbb{Z}^8 \backslash \{\vec{0}\}$. The resonant nature of the harmonic depends 
on the closeness of the actions $\lieI_0$ to the resonance condition 
\begin{equation}
\label{eq:res_condition}
\veck \cdot \lieomegamean_{2n}(\lieI) + \Omega = 0, \quad \lieomegamean_{2n} = \frac{\partial \liefourcoeff{0}{0}}{\partial \lieI}
,\end{equation}
which defines a hyperplane of normal vector $\veck$ in the space of the mean frequencies $\lieomegamean_{2n}$ and 
an algebraic variety in the space of the actions $\lieI$ (at degree four this is a hyperplane as $\lieomegamean_4$ depends 
linearly on $\lieI$). According to the reduced Hamiltonian, the actions $\lieI$ oscillate along 
the direction $\veck$, that is, 
\begin{equation}
\dot{\lieI} = -\frac{\partial h_{2n}^{\veck,\vecl}}{\partial \lietheta} = 
2 \veck \operatorname{Im} \{ \liefourcoeff{k}{\ell}(\lieI) \, \E^{j(\veck \cdot \lietheta + \Omega \tau)} \} 
.\end{equation}
Following the \citet{Chirikov1979} derivations, we performed the explicit reduction of the 
Hamiltonian in Eq. \eqref{eq:red_ham} to one DOF. We considered the canonical transformation 
$(\lieI,\lietheta) \rightarrow (\vecp,\vecvarphi)$ defined by the time-dependent generating function
\begin{equation}
F(\vecp,\lietheta,\tau) = \left( {\lieI_r}^T + \vecp^T \matM \right) \lietheta + \vecp^T \vec{\nu} \, \tau
,\end{equation}
where $\matM$ is a real $8 \times 8$ matrix, $\lieI_r, \vec{\nu} \in \mathbb{R}^8$, and $T$ denotes 
the transposition operator. One has
\begin{equation}
\begin{aligned}
&\lieI = \frac{\partial F}{\partial {\lietheta}^T} = \lieI_r + \matM^T \vecp,
\quad \vec{\varphi} = \frac{\partial F}{\partial \vecp^T} = \matM \lietheta + \vec{\nu} \tau, \\
&\hslash_{2n}^{\veck,\vecl} = h_{2n}^{\veck,\vecl} + \frac{\partial F}{\partial \tau} = h_{2n}^{\veck,\vecl} + \vecp^T \vec{\nu} .
\end{aligned}
\end{equation}
We chose the first row of the matrix $\matM$ to be $\veck^T$, and the remaining seven rows to be orthonormal vectors 
$(\vece^T_i)_{i=2}^8$ perpendicular to $\veck$, that is, $\vece_i^T \vece_\jmath = \delta_{i\jmath}$ and $\vece_i^T \veck = 0$ 
for $2 \leq i,\jmath \leq 8$. It follows that 
\begin{equation}
\lieI = \lieI_r + p_1 \veck + \sum_{i=2}^8 p_i \vece_i 
,\end{equation}
with $\vecp = (p_1, \dots, p_8)^T$. We finally chose $\vec{\nu} = (\Omega, 0, \dots, 0)^T$ so that 
$\varphi_1 = \veck \cdot \lietheta + \Omega \tau$. The reduced Hamiltonian $\hslash_{2n}^{\veck,\vecl}$ governing the new 
canonical variables $(\vecp, \vecvarphi)$ is given by 
\begin{equation}
\label{eq:red_ham_2}
\hslash_{2n}^{\veck,\vecl}(\vecp,\varphi_1) = 
\liefourcoeff{0}{0}(\lieI(\vecp)) + \Omega \, p_1 
+ 2 \operatorname{Re} \{ \liefourcoeff{k}{\ell}(\lieI(\vecp)) \, \E^{j \varphi_1} \}
.\end{equation}
The reduced dynamics only affects the momentum $p_1$, and $(p_i)_{i=2}^8$ are all integrals of motion. 
These constants can be set to zero by choosing $\lieI_r = \lieI_0$ and employing the initial condition 
$\lieI(\tau = t_0) = \lieI_0$ so that $p_1(\tau = t_0) = 0$ and 
\begin{equation}
\label{eq:action_motion_1}
\lieI(p_1) = \lieI_0 + p_1 \veck
.\end{equation}
As the action variables $\lieI$ are non-negative by definition, the dynamics of the momentum $p_1$ is generally 
restricted to a subset of the real line that is bounded from above, below, or both, depending on the wave vector $\veck$. 

\subsection{Reduced phase spaces}
A different choice considered by \citet{Chirikov1979} consists in taking $\lieI_r$ on the 
resonance surface given by Eq.~\eqref{eq:res_condition}, with the additional constraint that $\lieI_0 - \lieI_r$ 
is parallel to the direction $\veck$, that is, 
\begin{equation}
\label{eq:res_centre}
\veck \cdot \lieomegamean_{2n}(\lieI_r) + \Omega = 0, \quad \lieI_0 = \lieI_r + p_1(t_0) \veck
.\end{equation}
With this choice, the integrals of motion $(p_i)_{i=2}^8$ have null values and the action variables are given by 
\begin{equation}
\label{eq:action_motion}
\lieI(p_1) = \lieI_r + p_1 \veck
.\end{equation}
To simplify the notation, we omit the subscripts of the resonant variables $(p_1, \varphi_1)$ from now on. 
With Chirikov's choice and under the assumption of sufficiently small oscillations of the momentum $p$, 
the reduced Hamiltonian can be expanded around $\lieI_r$ and provides, at first order, the universal description 
of a non-linear resonance in terms of pendulum dynamics \citep{Chirikov1979}. In this case the vector $\lieI_r$ 
represents the centre of oscillation in the action space. 

When truncating the Lie transform in Eq. \eqref{eq:lie_transf} at degree $2n$, the two constraints in Eq.~\eqref{eq:res_centre} 
can be rewritten as a polynomial equation of degree $n-1$ in the variable $p(t_0)$, which thus possesses up to $n-1$ real 
solutions. As a generalisation of the pendulum approximation, when no real solutions for $p(t_0)$ exist, or when 
the corresponding vectors $\lieI_r$ have some negative components, generally no hyperbolic fixed points appear in the reduced 
phase space $(p,\varphi)$. The dynamics is thus globally characterised by rotation states of the angle $\varphi$. Depending on 
the Fourier amplitude $\liefourcoeff{k}{\ell}$, libration islands may still exist, 
for example close to the boundaries of the $p$ variable. However, in the absence of hyperbolic points, they are not enclosed by a separatrix. These reduced phase spaces are 
therefore considered as non-resonant. When a real solution for $p(t_0)$ exists and the corresponding point $\lieI_r$ lies 
in the physical action space (that is, its components are all non-negative), hyperbolic fixed points generally appear 
in the phase space. The separatrices that emerge from these points separate resonant libration states of $\varphi$ 
from rotation states. The position of the fixed points along the momentum axis is offset with respect to $p=0$ 
(corresponding to the action point $\lieI_r$) by an amount that depends on the Fourier coefficient $\liefourcoeff{k}{\ell}$. 

\subsection{Fixed points}
The topology of a reduced phase space depends on the position of the system in the eight-dimensional action space, 
that is, it depends on $\lieI_0$ through Eq.~\eqref{eq:action_motion_1} or Eq.~\eqref{eq:res_centre}. An explicit study 
of the one-DOF Hamiltonian in Eq. \eqref{eq:red_ham_2} can be carried out when a given point $\lieI_0$ is considered. 
The fundamental idea in this work is to retrieve the topology of the reduced phase spaces along the nominal solution 
$\mathcal{S}^\prime_{2n}(t) = (\lieI_{2n}(t), \lietheta_{2n}(t))$, that is, $\lieI_0 = \lieI_{2n}(t=t_0)$. For each Fourier harmonic 
$\veck,\vecl$, one obtains a one-parameter family of phase spaces spanned by the time $t_0$. We studied their topology 
through the retrieval of the fixed points. Their position can be analytically obtained in the pendulum approximation; 
however, while the pendulum model works fine for the majority of the harmonics, it turns out not to be pertinent 
for the leading resonant ones, which are our main interest. Fortunately, as the following derivations show, 
computer algebra allows the fixed points of a general reduced phase space to be systematically retrieved, making the 
pendulum approximation unnecessary. 

In light of Eqs.~\eqref{eq:action_motion_1} and \eqref{eq:action_motion}, the reduced Hamiltonian can be written as 
\begin{equation}
\label{eq:red_ham_3}
\hslash(p,\varphi) = 
f_0(\lieI(p)) + \Omega \, p
+ 2 \operatorname{Re} \{ f_1(\lieI(p)) \E^{j\varphi} \}
,\end{equation} 
where we rename $\hslash_{2n}^{\veck,\vecl}$, $\liefourcoeff{0}{0}$, and $\liefourcoeff{k}{\ell}$ as $\hslash$, $f_0$, and $f_1$, 
respectively, to keep a simpler notation. From Eq.~\eqref{eq:AA_vars} it follows that one can write 
$f_1(\lieI) = \mu(\!\sqrt{\lieI}) F_1(\lieI)$, with 
\begin{equation}
\label{eq:mu_factor}
\mu \left( \! \sqrt{\lieI} \right) = 
\prod_{i=1}^8 {\sqrt{\smash[b]{I^\prime_i}}}^{|k_i|} 
,\end{equation} 
where we denote $\sqrt{\lieI} = (\!\sqrt{\smash[b]{I^\prime_i}})_{i=1}^8$ and $\veck = (k_i)_{i=1}^8$. 
The function $\mu$ is thus a real multivariate monomial of degree $|\veck| = \sum_{i=1}^8 |k_i|$ in the 
square roots of the action variables, while $F_1$ can be expressed as a multivariate polynomial of degree $(2n - |\veck| - |\vec{\ell}|)/2$ 
in the action variables\footnote{The polynomial $F_1$ generally has complex coefficients as the phases of 
the quasi-periodic decomposition of the giant planet orbits in Eq.~\eqref{eq:qp_decomposition} are explicitly 
contained in the complex amplitudes.}, the order of the harmonic $\veck,\vecl$ (i.e. $|\veck| + |\vec{\ell}|$) being an even integer. 
The fixed points of the Hamiltonian $\hslash$ satisfy the system of equations 
\begin{equation}
\label{eq:fp}
\begin{cases}
\operatorname{Im}\{ f_1 \E^{j\varphi} \} = 0 \\
\dif f_0 + \Omega + 2\operatorname{Re}\{\dif f_1 \E^{j\varphi}\} = 0 
\end{cases} ,
\end{equation} 
where $\dif \,$ stands for derivation with respect to $p$. The function $\dif f_0$ is a multivariate 
polynomial of degree $n-1$ in the action variables and thus a univariate polynomial of the same degree 
when expressed in the variable $p$. Because of the square roots appearing in $f_1$ and in its derivative $\dif f_1$, 
the system of equations \eqref{eq:fp} is non-algebraic. Nevertheless, systematic manipulations of these equations 
allow its solutions to be found through the roots of a univariate polynomial in the variable $p$. 
Searching for solutions that correspond to non-null values of the action variables $\lieI$, we assumed $\mu \neq 0$ and 
write the first equation as $F_1 \E^{2j\varphi} = \conj{F_1}$. 
From Eq.~\eqref{eq:mu_factor} it follows that the square root dependence in $\dif f_1$ can be eliminated through 
multiplication by the function 
\begin{equation}
\label{eq:eta_factor}
\eta \left( \! \sqrt{\lieI} \right) = 
\prod_{i=1}^8 {\sqrt{\smash[b]{I^\prime_i}}}^{|k_i| \bmod{2}}
,\end{equation} 
that is, the function $\widetilde{\dif f_1} = \eta \dif f_1$ can be expressed as a polynomial 
of degree $n - 1 - (|\vec{\ell}|$ $ -\sum_{i=1}^8 |k_i| \bmod{2}) / 2$ in the action variables. 
Through multiplication by $\eta$ and taking the square, the second equation of system \eqref{eq:fp} implies 
$\eta^2 (\dif f_0 + \Omega)^2 = 4 \operatorname{Re}\{ \widetilde{\dif f_1} \E^{j\varphi} \}^2$. 
The first equation then allows the dependence on the variable $\varphi$  to be eliminated from the second equation, 
providing the system 
\begin{equation}
\label{eq:fp2}
\begin{cases}
F_1 \E^{2j\varphi} = \conj{F_1} \\
|F_1|^2 \left[\eta^2 (\dif f_0 + \Omega)^2 - 2|\widetilde{\dif f_1}|^2\right] 
- 2\operatorname{Re}\{(\widetilde{\dif f_1} \conj{F_1})^2\} = 0 
\end{cases} .
\end{equation} 
The second equation is now a univariate-polynomial equation in the variable $p$ of degree 
$2 (2n - 1 - \sum_{i=1}^8 \floor{|k_i|/2} ) - |\vec{\ell}|$. Its complex solutions can be systematically 
retrieved through a univariate polynomial solver. Among all the solutions, one has to keep the real ones that correspond 
to positive values of the action variables $\lieI$ in Eqs.~\eqref{eq:action_motion_1} or \eqref{eq:action_motion}. 
For each solution $p^\star$, the value of the angle $\varphi^\star$ is then given by $\E^{j\varphi^\star} = \pm (\conj{F_1}(p^\star)/F_1(p^\star))^{1/2}$, 
where the sign is chosen to restore the sign loss that occurs in squaring the second equation of the system \eqref{eq:fp}. 

The linear stability of a fixed point is assessed from the sign of the eigenvalues of the variational matrix 
\begin{equation}
\label{eq:stability}
\mathbb{V}  = 
\begin{pmatrix}
-\partial^2 \hslash / \partial p \partial \varphi & 
-\partial^2 \hslash / \partial \varphi^2 \\
 \partial^2 \hslash / \partial p^2 & 
 \partial^2 \hslash / \partial \varphi \partial p
\end{pmatrix}
\end{equation}
evaluated at the fixed point. The matrix $\mathbb{V}$ is the product of the symplectic matrix 
$\mathbb{J} =((0,-1),$ $(1,0))$ with the Hessian of the reduced Hamiltonian, and its eigenvalues $\lambda$ are given 
by the equation $\lambda^2 = - \det(\mathbb{V})$. In the case of an elliptic fixed point (i.e. when $\lambda^2 < 0$),
the angular frequency of the small oscillations is given by $\omegaell = \sqrt{\det(\mathbb{V})}$, while in the case of 
a hyperbolic fixed point (i.e. when $\lambda^2 > 0$) we denote $\omegahyp = \sqrt{-\det(\mathbb{V})}$. 

\subsection{Level curves}
We systematically characterised the level curves of the reduced Hamiltonian by retrieving their extrema. 
The tangent direction $\vec{t}$ to the level curve $\hslash(p,\varphi) = \hslash_0$ can be expressed as 
\begin{equation}
\label{eq:tang_vec}
\vec{t} = (\dot{p}, \dot{\varphi}) = (-\partial \hslash / \partial \varphi, \partial \hslash / \partial p)
.\end{equation}
The extrema of the variable $p$ along the level curve satisfy the equation $\partial \hslash / \partial \varphi = 0$, 
while the extrema of $\varphi$ are given by $\partial \hslash / \partial p = 0$. To distinguish between the minima and 
maxima of the variables, one studies the acceleration vector $\vec{a} = \dot{\vec{t}}$ along the level curve, which is given by 
\begin{equation}
\label{eq:acc_vec}
\vec{a} = \left( 
-\frac{\partial^2 \hslash}{\partial p \partial \varphi} \dot{p}
-\frac{\partial^2 \hslash}{\partial \varphi^2} \dot{\varphi} , 
\frac{\partial^2 \hslash}{\partial p^2} \dot{p} + 
\frac{\partial^2 \hslash}{\partial \varphi \partial p} \dot{\varphi}
\right)
.\end{equation}
It follows that $-(\partial \hslash / \partial p)(\partial^2 \hslash / \partial \varphi^2)$ is 
positive for a minimum of $p$ and negative in the case of a maximum. Similarly, an extremum of $\varphi$ is a minimum when 
$-(\partial \hslash / \partial \varphi)(\partial^2 \hslash / \partial p^2)$ is a positive quantity and a maximum 
if it is negative. 

As an example, the extrema of the momentum $p$ along the level curve $\hslash(p,\varphi) = \hslash_0$ 
are given by the system of equations 
\begin{equation}
\label{eq:extrema_p1}
\begin{cases}
\operatorname{Im}\{ f_1 \E^{j\varphi} \} = 0 \\
\hslash_0 = f_0 + \Omega \, p + 2 \operatorname{Re} \{ f_1 \E^{j\varphi} \}
\end{cases} .
\end{equation} 
Similarly to the computation of the fixed points, the solutions of Eq.~\eqref{eq:extrema_p1} can be searched for among those 
of the system of equations 
\begin{equation}
\label{eq:extrema_p1_2}
\begin{cases}
F_1 \E^{2j\varphi} = \conj{F_1} \\
(\hslash_0 - f_0 - \Omega \, p)^2 = 4 |f_1|^2 
\end{cases} ,
\end{equation} 
which involves the solution of a univariate-polynomial equation of degree $2n$ in the variable $p$. 

\subsection{Resonance half-width}
Following \citet{Chirikov1979}, we denote as $\omega$ the projection of the frequency vector $\lieomega = d\lietheta/dt$ 
along the normal to the resonance plane $\veck \cdot \lieomega + \Omega = 0$, that is, $\omega = \lVert\veck\rVert^{-1} \veck \cdot \lieomega$. 
The reduced dynamics in Eq. \eqref{eq:red_ham_3} induces a variation in this projection equal to $\halfwidth = \lVert\veck\rVert^{-1} \Delta \dot{\varphi}$. 
Based on ideas from frequency analysis \citep{Laskar1993}, we then associated with the level curve of the reduced Hamiltonian passing through 
the point $(p_0, \varphi_0)$ its rotational frequency $\nu(p_0, \varphi_0)$, that is, the time mean of the instantaneous frequency $\dot{\varphi}$,
\begin{equation}
\label{eq:rot_freq}
\nu(p_0, \varphi_0) = \lim_{t \to \infty} \frac{1}{t} \int_0^t \dot{\varphi}(\tau; p_0, \varphi_0) \, d\tau
,\end{equation}
where the time integration is meant along the trajectory starting at $(p_0, \varphi_0)$ at time $\tau = 0$. 
The rotational frequency of a separatrix is null, as for the librations of the angle $\varphi$ inside a resonant island, 
while $\nu$ is different from zero for rotations. Therefore, we established the half-width of a resonant 
harmonic from the variation in the rotational frequency across its separatrix. To associate an intrinsic 
half-width with each harmonic as in the \citet{Chirikov1979} definition, we defined once and for all 
the two-fold half-width as 
\begin{equation}
\label{eq:half_width_pm}
\halfwidth_\pm = \lVert\veck\rVert^{-1} \nu(p_\pm, \varphi_\pm + \pi/2) 
,\end{equation}
where $(p_\pm, \varphi_\pm)$ are the locations of the maximum and minimum, respectively, of the momentum $p$ 
along the separatrix of the resonance (it is implicitly assumed that the line curve through $(p_\pm, \varphi_\pm + \pi/2)$ corresponds 
to a rotation of the angle $\varphi$). When applied to the simple pendulum, definition \eqref{eq:half_width_pm} gives 
\begin{equation}
\label{eq:half_width_pendulum}
\halfwidth_\pm = \pm 2 \omegahyp \lVert\veck\rVert^{-1} \beta
,\end{equation}
where $\beta = \pi / (2 \sqrt{2/3} \mathcal{K}(2/3))$ is a constant factor, $\mathcal{K}(m) = \int_0^{\pi/2} (1 - m \sin^2 \phi)^{-1/2} d\phi$ 
being the complete elliptic integral of the first kind. As $\beta \approx 0.95$, Eq.~\eqref{eq:half_width_pendulum} is practically the 
\citet{Chirikov1979} half-width (recall that $\omegaell = \omegahyp$ for the pendulum). 

To compute the frequency $\nu$ of a rotating line curve in a numerically efficient way, we write 
$\nu(p_0, \varphi_0) = 2\pi/\mathcal{T}(p_0, \varphi_0)$, where the signed period $\mathcal{T}$ is given by 
\begin{equation}
\label{eq:signed_period}
\mathcal{T}(p_0, \varphi_0) = \int_0^{2\pi} \frac{d\varphi}{\dot{\varphi}(p(\varphi; p_0, \varphi_0), \varphi)} ,
\end{equation}
with $\dot{\varphi}(p,\varphi) = \partial \hslash(p,\varphi)/\partial p$. Equation \eqref{eq:signed_period} assumes 
that $\varphi$ is a monotonic function of time along the level curve through $(p_0, \varphi_0)$ (i.e. $\dot{\varphi} > 0$ or $\dot{\varphi} < 0$), and $p = p(\varphi; p_0, \varphi_0)$ then represents the momentum 
as a (single-valued) function of the angle\footnote{In the general case, the integral must be split over 
subsets of $[0,2\pi]$ for which $\dot{\varphi}$ does not change sign, and the contributions summed in absolute value to obtain 
the period of the motion. We simply do not compute the rotational frequency when $\varphi$ is not monotonic, 
as this constitutes a minority of cases.}. 
To compute the signed period without numerical integration of the corresponding trajectory, we reconstructed the curve $p = p(\varphi; p_0, \varphi_0)$ 
geometrically. We started by retrieving the minimum, $p_\textrm{min}$, and maximum, $p_\textrm{max}$, of the momentum along the 
level line $(p_0, \varphi_0)$ as solutions of system \eqref{eq:extrema_p1_2}. We then linearly sampled values of $p$ 
in the interval $[p_\textrm{min},p_\textrm{max}]$ and retrieved corresponding values for $\varphi$ through Eq.~\eqref{eq:red_ham_3}, 
which consists of a linear trigonometric equation in the angle variable. We finally used spline interpolation 
to uniformly sample the curve along the angle axis and compute the integral in Eq. \eqref{eq:signed_period}. 

The two-fold definition in Eq. \eqref{eq:half_width_pm} attributes a different half-width to each side of the resonance 
along the momentum axis. While the pendulum dynamics is symmetric, the two half-widths $\halfwidth_\pm$ are not equal 
in absolute value in the general case. Moreover, their signs depend on the topology of the reduced phase space and 
are not necessarily opposite (see, as an example, the resonance $\theta_{1:1}$ at 450 Myr in Fig.~\ref{fig:phase_space_1}). 
The time statistic of the resonance half-width along the sampled nominal solution (given in Table~\ref{tab:resonant_harmonics}) 
is constructed from the sample 
\begin{equation}
\label{eq:half_width}
\halfwidth = \{ |\halfwidth_+(t_m)|, |\halfwidth_-(t_m)| \}
,\end{equation}
where $m$ spans the subset of times the harmonic is resonant. The two sides of the resonance are thus equally weighted in absolute 
value. More informative statistics are the signed half-widths, $\halfwidth^+$ and $\halfwidth^-$, derived from the samples 
\begin{equation}
\label{eq:half_width_signed}
\begin{cases}
\halfwidth^+ = \{ \halfwidth_\pm(t_m) \, | \, \halfwidth_\pm(t_m) > 0 \} \\
\halfwidth^- = \{ \halfwidth_\pm(t_m) \, | \, \halfwidth_\pm(t_m) < 0 \}
\end{cases} ,
\end{equation}
whose medians define the asymmetric resonance layers in Fig.~\ref{fig:overlaps}. 

\subsection{Time sampling}
To retrieve the samples of the fixed point frequencies $\omegahyp$, $\omegaell$ and of the resonance half-width $\halfwidth$ 
along the sampled nominal solution $\mathcal{S}^\prime_{2n}(t_m) = (\lieI_{2n}(t_m), \lietheta_{2n}(t_m))$, we first applied the 
low-pass KZ filter to the time series of the action-angle variables, 
\begin{equation}
(\widehat{\lieI}_{2n}(t_m), \widehat{\lietheta}_{2n}(t_m)) = \textrm{KZ}[(\lieI_{2n}(t_m), \lietheta_{2n}(t_m))] 
,\end{equation}
with three iterations of the moving average and a cutoff frequency of (5 Myr)$^{-1}$. Similarly to our definition of a librating 
harmonic in Appendix \ref{supp:librating_harm}, we are not interested in harmonics that are resonant over timescales much shorter 
than the Lyapunov time. Moreover, low-pass filtering allows the time series to be resampled with a timestep $\Delta t^\prime$ = 250 kyr, which is 
much bigger than the original one, $\Delta t$ = 1 kyr \suppref{supp:nomsol}. This resampling is fundamental for actually being able 
to perform the systematic retrieval of the reduced phase-space topology over the entire nominal solution, which spans 5 Gyr, 
and for all the librating harmonics. We point out that, because of the linearity of the KZ filter, filtering the angle variables 
$\lietheta(t)$ is equivalent to filtering the corresponding instantaneous frequencies $d\lietheta(t)/dt$, as done in Appendix \ref{supp:librating_harm} 
to define time-dependent fundamental frequencies. 

\subsection{Polynomial solver}
To find numerical approximations to the roots of a univariate polynomial with complex coefficients, TRIP implements 
the fixed-precision algorithm of \citet{Bini1996} and extends it to quadruple- and multi-precision floating point coefficients. 
Based on the Ehrlich-Aberth iteration \citep[e.g.][]{Bini1996}, which allows simultaneous approximations of all the roots, the algorithm
guarantees a posteriori error bounds. This allows, in particular, the real roots 
we are interested in to be isolated in a robust way. The algorithm deals with polynomials of high degree and with 
very large or small coefficients. Such polynomials can be generated in the symbolic preprocessing of systems 
of equations, as in our case. When a lack of convergence to some of the roots is detected 
after a fixed maximum number of iterations, or in the rare case of an overflow, we temporarily switched 
to the more recent implementation of MPSolve~3 \citep{Bini2000,Bini2014}, which we integrated in TRIP and 
which provides a guaranteed approximation of the roots with any desired number of digits. 

\begin{figure*}
\includegraphics[width=2\columnwidth]{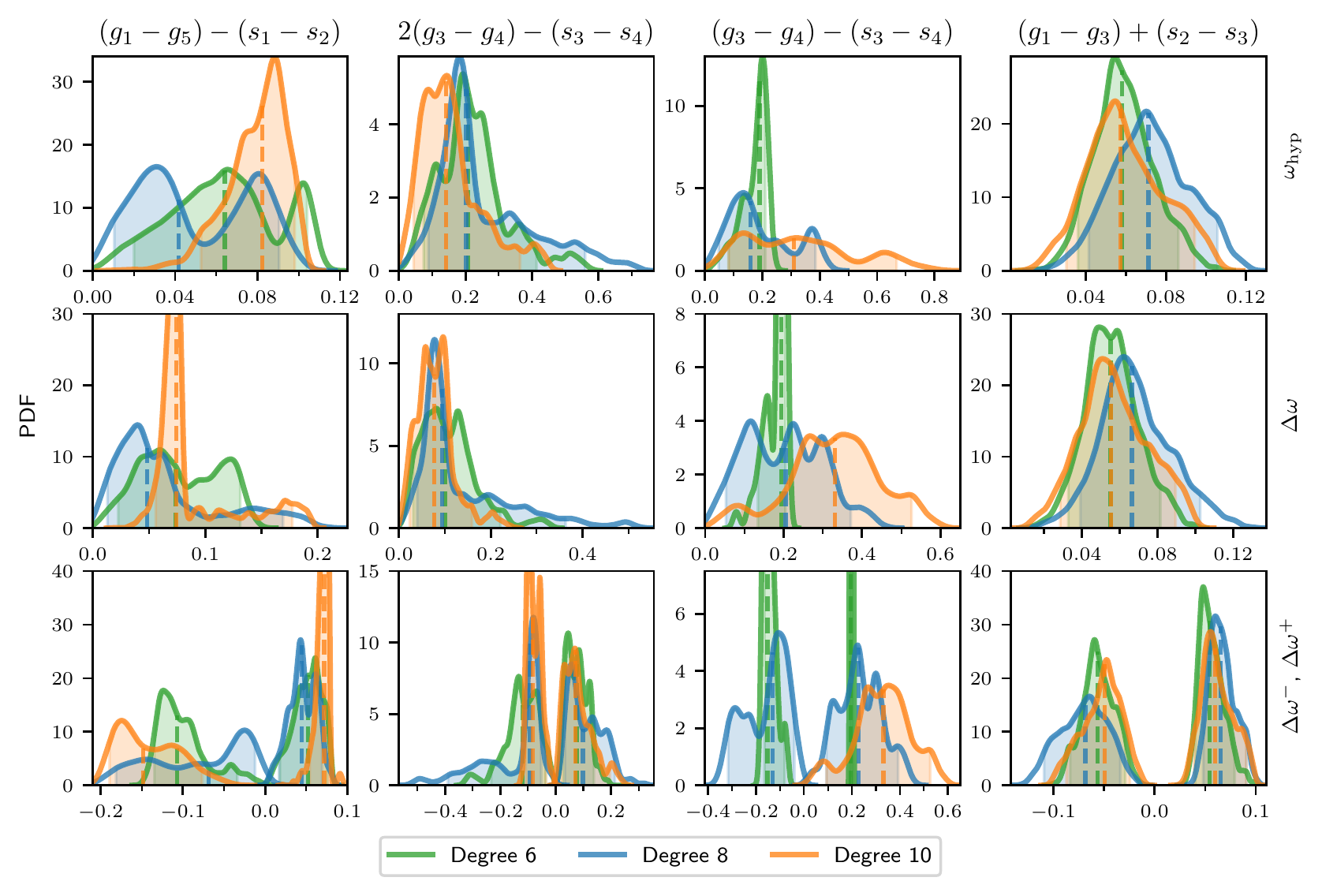}
\caption{Time PDF of the hyperbolic fixed point frequency $\omegahyp$ (first line), resonance half-width $\halfwidth$ 
(second line), and signed half-widths $\halfwidth^-$, $\halfwidth^+$ (third line), for four resonant harmonics 
listed in Table~\ref{tab:resonant_harmonics} and degree of truncation $2n$ from six to ten. Each time distribution 
results from the truncated Hamiltonian $\lieham_{2n}$ along the nominal solution $\mathcal{S}^\prime_{2n}(t)$ spanning 5 Gyr. 
Kernel density estimation of the PDF is shown as a solid line, and the observed median and the [5$^\textrm{th}$, 95$^\textrm{th}$] 
percentile range are represented by a vertical dashed line and a shaded region, respectively. Frequencies are in \arcsecyrtext{}.} 
\label{fig:trunc_modes}
\end{figure*}

\section{Transformed proper modes and time distributions}
\label{supp:transf_proper_modes}
The proper modes $\vecu = (u_i)_{i=1}^4$ and $\vecv = (v_i)_{i=1}^4$ introduced in Appendix \ref{supp:model} change under the Lie transform according to 
\begin{equation}
\label{eq:transf_proper_modes}
(\lieu, \liev) = \E^{-L_\mathrm{S}} (\vecu, \vecv)
.\end{equation}
While the transformed proper modes are formally determined by Eq.~\eqref{eq:transf_proper_modes}, the need to truncate 
the Lie transform in actual computations produces different definitions, that is, 
\begin{equation}
\label{eq:trunc_proper_modes}
(\lieu_{2n}, \liev_{2n}) = \left( \sum_{q=0}^{n-1} \frac{L_\mathrm{S}^q}{q!} \right) (\vecu, \vecv)
.\end{equation}
Even though the Lie transform defines a canonical change of variables close to identity, its slow (finite-degree) convergence 
shown in Appendix \ref{supp:lie} suggests that the contributions from high-degree terms are actually non-negligible. 
Indeed, harmonics of order eight and ten appear among the leading resonances of Table~\ref{tab:resonant_harmonics}, and their amplitudes 
are largely the result of the Lie transform. 
The truncated proper modes in Eq.~\eqref{eq:trunc_proper_modes} may thus be expected to differ from one another in an 
appreciable way. As the resonances in Table~\ref{tab:resonant_harmonics} are given in the proper modes truncated at degree 
ten, we show in Fig.~\ref{fig:trunc_modes} some examples of the dependence of the time statistic on the degree of truncation 
of the Lie transform. For the resonant harmonics $\sigma_{1:1}$, $\theta_{2:1}$, $\theta_{1:1}$, and $(g_1-g_3)+(s_2-s_3)$, 
we report in the corresponding column the time PDF of the hyperbolic fixed point 
frequency $\omegahyp$ (first line), the resonance half-width $\halfwidth$ (second line), and the signed half-widths 
$\halfwidth^-$, $\halfwidth^+$ (third line), for degrees of truncation from six to ten. Each time distribution thus results 
from the truncated Hamiltonian $\lieham_{2n}$ along the corresponding nominal solution $\mathcal{S}^\prime_{2n}(t),$ which spans 5 Gyr. 
We show the kernel density estimation \citep{Rosenblatt1956,Parzen1962} of each PDF (we used the Gaussian kernel 
and the \citet{Silverman1986} rule of thumb to select the bandwidth\footnote{The harmonic $\theta_{1:1}$ is 
resonant for only 0.3\% of the time at degree six, so the corresponding PDFs present spikes. We thus had to adapt the bandwidth choice 
to get a smoother estimation for $\halfwidth$. We also point out that $\theta_{1:1}$ possesses no negative signed half-width 
at degree ten (see its phase space at 450 Myr in Fig.~\ref{fig:phase_space_1}).}) as well as the observed median 
and the [5$^\textrm{th}$, 95$^\textrm{th}$] percentile range. 
Figure \ref{fig:trunc_modes} shows that the time distributions of different degrees largely overlap one another. Their medians, 
in particular, differ by a factor of two at most. These examples indicate that, as the nominal solution varies within the chaotic zone, 
the resonant nature of a Fourier harmonic can be stated statistically, in a way that is, within the framework of this study, 
largely independent of the precise truncation of the proper modes in Eq.~\eqref{eq:trunc_proper_modes}. 
The dependence on the truncation degree does not modify the general picture of the resonance web given 
in Table~\ref{tab:resonant_harmonics}. Truncation at degree ten, used for that table, 
allows the appearance of higher-order harmonics  to be accounted for. In this context, we point out that one may expect 
libration episodes of a resonant harmonic to emerge in a statistical framework independently of the precise proper 
modes and even the dynamical model employed, as shown in Fig.~\ref{fig:librations}. This consideration 
is clearly restricted to sets of proper modes that differ by a quasi-identity change of variables from those that integrate 
the LL dynamics (e.g. $\vecu$, $\vecv$ in this work or $\vec{z}^\bullet$, $\vec{\zeta}^\bullet$ in \citealt{Laskar1990}), 
and to dynamical models that faithfully reproduce the predictions of $N$-body numerical integrations. 

We remark that we do not provide confidence intervals for the PDF estimation in Fig.~\ref{fig:trunc_modes}. Estimation of 
confidence intervals must take into account that the time samples are correlated over a length comparable to the Lyapunov time of the dynamics. 
Confidence intervals for correlated data can be obtained via moving-block bootstrap \citep{Hoang2021}, for example. While one may expect the variance of the PDF estimation to be rather large, we aim here to show the compatibility of the distributions across different degrees 
of truncation, and this is already indicated by the PDF estimation itself. More precisely, we do not intend to state exact time distributions 
for $\omegahyp$ or $\halfwidth$ (which would be limited by the assumptions behind our approach, by the way) but simply to identify 
the ranges over which these dynamical quantities may statistically vary. As an example, the overlap between the asymptotic Lyapunov exponent 
in Fig.~\ref{fig:lyap} and the ranges spanned by $\omegahyp$ and $\halfwidth$ in the case of the leading resonances $\theta_{1:1}$ and $\theta_{2:1}$ 
is clearly robust across different degrees of truncation. In this sense, the medians and percentile ranges in Table~\ref{tab:resonant_harmonics} 
are not meant to represent strict values for the corresponding observables, but more simply to synthesise distributions such as those shown 
in Fig.~\ref{fig:trunc_modes}. 

\end{appendix}

\end{document}